\documentclass[12pt,a4paper]{article}
\usepackage[utf8]{inputenc}
\usepackage{amsmath}
\usepackage{enumerate}
\usepackage{amsfonts}
\usepackage{amssymb}
\usepackage{mathrsfs}
\usepackage{graphicx}
\usepackage{cite}
\usepackage{latexsym}
\usepackage{braket}
\begin{document}
\vskip 2cm

\def\warning#1{\begin{center}
\framebox{\parbox{0.8\columnwidth}{\large\bf #1}}
\end{center}}

\begin{center}

{\large {\bf $(3+1)$-Dimensional Topologically Massive 2-form Gauge Theory: Geometrical Superfield Approach}}

\vskip 2.0 cm

{\sf{ \bf R. Kumar$^1$ and Debmalya Mukhopadhyay$^{2,3}$}}\\
\vskip .1cm
{\it $^1$Department of Physics \& Astrophysics,\\ University of Delhi, New Delhi--110007, India}\\
\vskip .1cm
{\it $^2$Department of Theoretical Physics,\\ Indian Association for the Cultivation of Science,\\ 
2A \& B Raja S.C. Mullick Road, Jadavpur, Kolkata--700032, India}\\
\vskip .1cm
{\it $^3$Present Address: Variable Energy Cyclotron Centre  (VECC), 1/AF, Bidhannagar, Kolkata,  West Bengal--700064, India}\\
\vskip .2cm
{\tt{E-mails: raviphynuc@gmail.com; debphys.qft@gmail.com  }}\\
\end{center}

\vskip 1.5 cm

\noindent

\noindent
{\bf Abstract:} We derive the complete set of off-shell nilpotent and absolutely anticommuting Becchi--Rouet--Stora--Tyutin (BRST)
 and anti-BRST symmetry transformations corresponding to the combined ``scalar" and ``vector" gauge symmetry transformations for 
the $(3+1)$-dimensional (4D) topologically massive non-Abelian $(B \wedge F)$ theory with the help of geometrical 
superfield formalism. For this purpose, we use {\it three} horizontality conditions (HCs). The first HC produces the 
(anti-)BRST transformations for the 1-form gauge field and corresponding (anti-)ghost fields whereas the second HC yields the 
(anti-)BRST transformations for 2-form field and associated (anti-)ghost fields. The integrability of second HC produces third HC. 
The latter HC produces the (anti-)BRST symmetry transformations for the compensating auxiliary  vector field and corresponding ghosts. 
We obtain {\it five} (anti-)BRST invariant Curci--Ferrari (CF)-type conditions which emerge very naturally as the 
off-shoots of superfield formalism. Out of five CF-type conditions, two are fermionic in nature.  These CF-type conditions  
play a decisive role in providing the absolute anticommutativity of the (anti-)BRST transformations and also responsible 
for the derivation of coupled but equivalent (anti-)BRST invariant Lagrangian densities. Furthermore, we capture the 
(anti-)BRST invariance of the coupled Lagrangian densities in terms of the superfields and translation generators along 
the Grassmannian directions $\theta$ and  $\bar \theta$.    
    
\vskip 1.5 cm

\noindent



\newpage


\section{Introduction}
Every $p$-form  ($p = 1,2,3,...$) gauge theory remains invariant under a global symmetry known as BRST symmetry when 
we include the gauge-fixing term  and Faddeev--Popov ghosts in the theory~\cite{sw:1996,becchi:1975a,becchi:1975b,becchi:1975c,Tyutin:1975}. 
The physical significance 
of the BRST symmetry is to provide the unitarity in various interactions under consideration~\cite{kugojima}. The BRST symmetry 
transformation of the fields is generated by the BRST charge $Q_b$ which is nilpotent $\left(Q^2_b=0\right)$. The nilpotency 
leads to the formation of BRST cohomology where the physical state $|phys \rangle$, defined by $Q_b |phys \rangle=0$, 
is equivalent to another physical state $|phys \rangle'$ if $|phys \rangle'=|phys \rangle + Q_b|phys \rangle$. 
From this equivalence, we can identify the unphysical modes of a state (in the total quantum Hilbert space of states)  whose contributions are 
mutually cancelled  in a physical process~\cite{kugojima}. Consequently, the unitarity is achieved  in a given physical process.

The BRST formalism is one of the most elegant and mathematically rich methods to covariantly quantize any arbitrary $p$-form  
(non-)Abelian gauge theory. For a  given {\it classical} gauge symmetry, we have two linearly independent global 
supersymmetric type {\it quantum} BRST and anti-BRST symmetries. The latter symmetries are nilpotent of order two (i.e. $s^2_b = 0,\; s^2_{ab} = 0$) 
and absolutely anticommuting (i.e. $s_b s_{ab} + s_{ab} s_b = 0$) in nature \cite{cur,oji}. The absolute anticommutativity property 
of the (anti-)BRST transformations for the non-Abelian 1-form gauge theory and higher form ($p \geq 2$) (non-)Abelian gauge theories 
is satisfied due to the existence of Curci--Ferrari (CF)-type conditions \cite{cur,rpm1,rpm2,rpm3}. Furthermore, the CF-type conditions 
also play an important role in the derivation of coupled (but equivalent) Lagrangian densities.  These CF-type conditions emerge automatically 
within the framework of superfield formalism \cite{rpm3,bt1,rpm4}. The emergence of CF-type of condition(s) is one of the characteristic features of 
a $p$-form (non-)Abelian gauge theory within the framework of superfield approach to BRST formalism.

Bonora--Tonin  superfield approach to BRST formalism is a geometrical method to derive the proper off-shell nilpotent and 
absolutely anticommuting (anti-)BRST symmetry transformations for a given gauge theory \cite{bt1,bt2,del}. In this formalism, we generalize an
ordinary D-dimensional Minkowskian space to the (D, 2)-dimensional superspace with the help of a pair of Grassmannian coordinates ($\theta, \bar \theta$)
(with $\theta^2 = \bar \theta^2 = 0, \; \theta \bar \theta + \bar \theta \theta = 0$)
in addition to the ordinary bosonic coordinates $x^\mu $ ($\mu = 0, 1, 2, 3, ..., D - 1$). Further, we generalize the dynamical fields to
their corresponding superfields onto the (D, 2)-dimensional supermanifold. By exploiting the power and strength of celebrated horizontality 
condition (HC) \cite{bt1,bt2,del,mieg1,mieg2,hoyo1,hoyo2,cotta,balu,manes,bonora}, we 
obtain the desired (anti-)BRST symmetry transformations. 
The HC implies that the components of super curvature along the Grassmannian 
directions are zero. Physically, the HC condition demands that the gauge-invariant quantities should be
independent of the Grassmannian coordinates. In other words, the gauge-invariant quantities should not be affected by the presence of 
Grassmannian variables when they are generalized on the supermanifold.

The HC carries a very important physical significance in the gauge theories.  
Since, Faddeev-Popov-DeWitt ghost fields belongs to L(G), where $L(G)$ is a set of left-invariant one-forms being always isomorphic to tangent space at identity on group manifold $T_e(G)$~\cite{mann}, then HC implies that the equivalent representation of gauge field is always connected to the identity i.e. there is no anomaly due to BRST transformation in gauge theory without the matter fields~\cite{mann}. This conclusion can be similarly drawn from HC for Kalb-Ramond field. As a consequence, there is no anomaly of color current if the topologically massive model is applied in QCD without matter fields. Anomaly may be present when the model contains massless fermions with suitable action. In that case,  Wess and Zumino found a consistent condition to be obeyed if quantum action of matter content gauge theory is not gauge invariant~\cite{sw:1996,mann,wess,bilal}. In that case, HC leads to provide Wess-Zumino consistency condition for anomaly or Stora-Zumino chain of descent equations~\cite{manes}.

In recent years, the ``augmented" superfield formalism (which is an extended version of 
Bonora--Tonin superfield approach) has been extensively used for the interacting
gauge theories such as 1-form gauge theory interacts with Dirac's fields and complex scalar fields \cite{rpm7}, gauge-invariant 
Proca theory \cite{rpm8}, gauge-invariant massive 2-form theory \cite{rk1} and references therein. In this approach, in addition to the HC, the 
conserved currents and/or gauge-invariant restrictions play very important role in the derivation of the complete set of (anti-)BRST  transformations.

During last few decades, the antisymmetric Kalb--Ramond field $B_{\mu\nu} (= - B_{\nu\mu})$ of rank two became quite popular because of its relevance in the 
context of (super-)string theories \cite{gre,pol}, (super-)gravity theories \cite{sal}, dual description of a massless scalar field \cite{des,aur} and  
noncommutative theories \cite{sei}. It has been shown, within the framework of BRST formalism, that the 4D free Abelian 2-form gauge theory provides a 
tractable field-theoretic model for Hodge theory where de Rham cohomological operators of differential geometry and Hodge duality 
operation find their physical realizations in terms of the continuous and discrete symmetries, respectively \cite{rpm9}. Furthermore, it has also shown 
to be a quasi-topological field theory (q-TFT) which captures some features of Witten-type TFT and some aspects of Schwartz-type TFT \cite{rpm10}.

The 2-form antisymmetric gauge field also plays an important role in the mass generation of the vector gauge bosons through 
a well-known topological $(B \wedge F)$ term~\cite{crem,lah1,lah2,mieg,oda,lah3,lah4}. In this model, the mass of gauge bosons
 and gauge-invariance co-exist together. 
The phenomenological aspects of this model have been discussed in~\cite{deb} which shed light on the various kind of physical processes that are 
allowed by the standard model of particle physics. We have also studied the 4D (non-)Abelian topologically massive theory within the 
framework of BRST formalism~\cite{rk2,rk3}. In earlier work~\cite{rpm2}, the 4D non-Abelian topologically massive gauge theory has 
been studied in the context of superfield formalism where the ``scalar" and ``vector" gauge symmetries have been treated separately.  
As a consequence, the (anti-)BRST symmetry transformations
corresponding to the above gauge symmetries are found to be off-shell nilpotent and absolutely anticommuting. We point out that 
when  we combine the (anti-)BRST transformations corresponding to the scalar and vector gauge transformations, the resulting 
(anti-)BRST transformations are found to be off-shell nilpotent but they do not obey the absolute anticommutativity property. 
In our present investigation, we shall investigate this issue and
 derive the proper (anti-)BRST symmetries for the combined scalar and vector gauge transformations.

We know that pure Yang-Mills (YM) theory~\cite{ym:1954} obeys unitarity where the 1-form gauge field is taken to be massless. 
Due to having mass, the 1-form gauge field has a physical longitudinal mode. 
But the scattering among the longitudinal modes shows the violation of unitarity in tree level scattering processes
~\cite{bell:1972,Smith:1973}. We know this happens when we consider the tree level $2\to2 $ scatterings of longitudinally polarized massive 
gauge bosons ($W^{\pm}$ and $Z^0$) in electroweak sector of the standard model excluding the process mediated by 
Higgs particle~\cite{jog:1973,lee:1977}. These are the Higgs mediated processes which save the unitarity of the 
scattering process among the longitudinally polarized electroweak bosons. In the Higgs 
mechanism~\cite{higgs:1964,englert:1964}, global symmetry $SU(2)_L\times U(1)_Y$ is spontaneously broken to the 
electromagnetic $U(1)$ group~\cite{SW:1967,AS:1968,glashow:1961,sw:1995}. But we will consider the 4D dynamical $(B \wedge F)$ theory where mass 
of gauge boson is generated and keeping the global $SU(N)$ symmetry unbroken.

Using the geometric features of a gauge theory~\cite{mieg1,daniel:1980,bleecker,spivak}, we obtain {\it proper} (anti-)BRST transformations
for all the fields. But it is not guaranteed  
whether the BRST charge keeps its nilpotency after quantum corrections. The assurance of unitarity at every order of quantum correction 
comes from the renormalizability of model. Pure YM theory containing massless 1-form gauge field is an example where the BRST symmetry 
and renormalizibility are maintained simultaneously. The mass generation of  YM field (keeping global symmetry unbroken) shows
 unsatisfactory characteristics in quantum field theory. For example, non-Abelian St{\"u}ckelberg model is found to be 
non-renormalizable~\cite{umezawa:1961,veltman:1968-1970a,veltman:1968-1970b,sizuya:1976,ruegg:2004} but it obeys unitarity. 
On the other hand, Curci--Ferrari model~\cite{curci:1975} containing Proca massive non-Abelian YM field  shows renormalizibility 
but it fails unitary in $(3+1)$-dimensions~\cite{ojima:1982,boer:1996}. There is a possibility of the mass generation by 
dynamical symmetry breaking in non-perturbative regime, but the mass tends to zero at the high energy limit of non-Abelian 
gauge theory~\cite{cornwall:1982}. The BRST symmetry plays an important part in the analysis of the various interactions 
according to a model under consideration. We should need the unitarity of the scattering matrix (S-matrix) in a renormalizable model.

Our present investigation is essential on the following grounds.  First, to derive the proper off-shell nilpotent and absolutely anticommuting 
(anti-)BRST transformations for combined scalar and vector gauge transformations. Because in earlier work \cite{rpm2}, the off-shell nilpotent 
(anti-)BRST transformations are found to be non-anticommuting. Second, to obtain the coupled and equivalent Lagrangian densities which
respect both BRST and anti-BRST transformations. Third, to establish the CF-type of conditions because these conditions play an important role 
within the framework of BRST formalism.


The contents of our present endeavour are organized as follows. In section 2,
we briefly  discuss about the mathematical aspects and geometrical significance of the BRST symmetries 
in the realm of differential geometry. In section 3, we discuss about the 4D topologically 
massive  (non)-Abelian $(B \wedge F)$ theories and associated local gauge symmetries. 
Section 4 deals with the derivation of the {\it proper} off-shell nilpotent and absolutely anticommuting (anti-)BRST symmetry transformations of the 
Yang-Mills field, antisymmetric gauge field and compensating auxiliary vector field and their corresponding 
(anti-)ghost fields within the framework of geometrical superfield approach to BRST formalism. 
Section 5 is devoted to the derivation of the coupled (but equivalent) Lagrangian densities by using the basic tenets of BRST formalism. 
We capture, in section 6, the (anti-)BRST invariance of the coupled Lagrangian densities, nilpotency and absolute anticommutativity properties of the 
(anti-)BRST symmetries within the framework of superfield formalism. 
Finally, in section 7, we provide some concluding remarks.

In our Appendix {\bf A}, we show the precise values of the various secondary field that are presented in the supetfield expansions  
in terms of the dynamical and auxiliary fields of the (anti-)BRST invariant theory. Appendix {\bf B} deals with the proof of the 
absolute anticommutativity of the (anti-)BRST transformations where the CF-type of conditions play decisive role. 
The (anti-)BRST invariance of the coupled Lagrangian densities is
shown in Appendix {\bf C}.

\section{Geometrical significance of BRST symmetries: mathematical aspects}

In this section, we consider the geometrical significance of the BRST symmetry 
(see, e.g.~\cite{mieg1,daniel:1980,bleecker,spivak} for details). 
We need to consider the principal $G$-bundle ($P$, $\pi$, $M$) in pure YM theory 
\begin{figure}[h!]
\begin{center}
\includegraphics[scale=0.04]{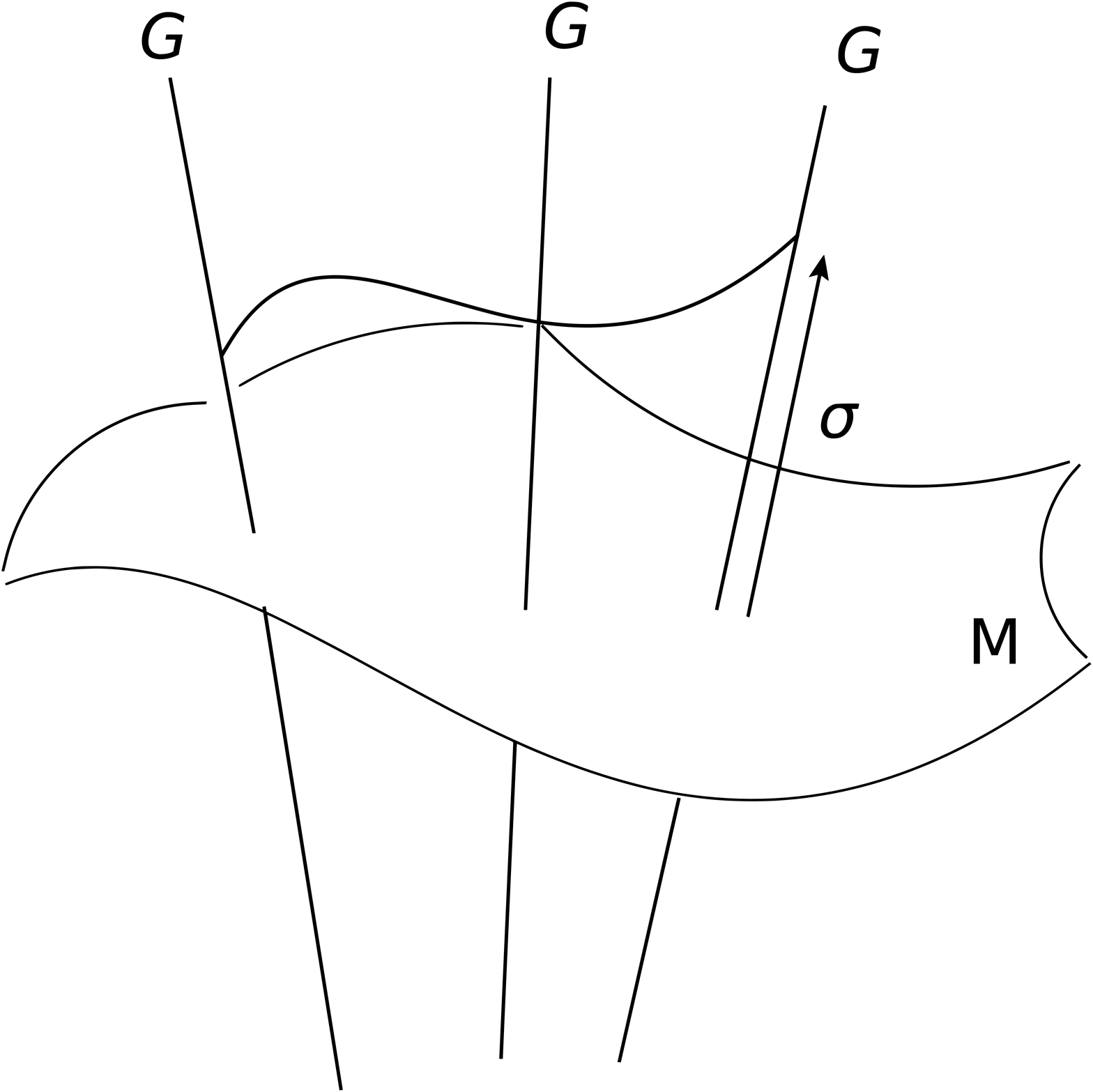}
\caption{Fibres $G$ in the principal  fibre bundle with base manifold $M$ and section $\sigma$}
\label{fig1}
 \end{center} 
\end{figure} 
where $F \equiv G$ is the fibre  in the total space $P$ and $G$ is the structural Lie group over the base manifold which is  
spacetime (see Fig. 1). Here $\pi$ is the projection of $F$ on the $M$. We define a section $\sigma :M\to F$ such that 
\begin{eqnarray}
\pi(\sigma(y))=id_M, \quad y\in G,
\label{section}
\end{eqnarray}
and Lie algebra valued connection 1-form $\omega$ on the bundle. The pull-back  $\omega$ on $M$ i.e. 
$\sigma^*\omega$ represents YM field locally (or local trivialization). Here $id_M$ in the Eq. (\ref{section}) 
represents an identity map of $M$. Let us consider the coordinates $y^i$ in the fibre and a point $x^\mu$ on $G$ 
which is lifted  in $\sigma$ from $M$. 
The vector $\partial_{y^i}$ is tangent to the fibre and vertical whereas the vector $\partial_{x^\mu}$ is tangent to the section but
neither horizontal nor vertical. The 1-forms $dy^i$ and $dx^\mu$ span the cotangent space $P^*$. Thus, 1-form $\omega$ can be decomposed as
\begin{equation}
\omega=\chi_i dy^i+\phi_\mu dx^\mu,
\label{decompose}
\end{equation}
where $\chi = \chi_i dy^i$ is the Maurer--Cartan form, which is Faddeev-Popov ghost field on the bundle and $\phi_\mu$ is the 
1-form  gauge field. The ghost field $\chi$ is vertical and  defined as 
\begin{eqnarray}
\chi_i(\partial_{x^\mu})=0,
\label{vertical}
\end{eqnarray}
whereas the gauge field is horizontal:
\begin{eqnarray}
\phi_\mu(\partial_{y^i})=0.
\label{horizontal}
\end{eqnarray}
We can also decompose the exterior derivative $d$ of a $0$-form according to the Eq. (\ref{decompose}) as 
\begin{eqnarray}
d f = s f + b f,
\end{eqnarray} 
where $s$ and $b$ are defined in the following fashion:
\begin{eqnarray}
sf=\partial_{y^i} f dy^i, \qquad b=\partial_{x^\mu}f \,dx^\mu.
\end{eqnarray}
Using the cohomology with respect to exterior derivative, we obtain
\begin{eqnarray}
s^2 = 0, \qquad b^2 = 0, \qquad sb + bs = 0,
\end{eqnarray} 
In the above, 
$s$ defines the exterior differential normal to the sections and it is nilpotent of order two
whereas $b$ is horizontal operator.  We shall identify  $s$ as the BRST operator.

Due to the construction of the fibre  bundle, we can clearly see
\begin{eqnarray}
\sigma^*(df)=\sigma^*(bf),
\end{eqnarray}
because 
\begin{eqnarray}
\sigma^*(sf)=0.
\end{eqnarray}
Then the 2-form curvature  with respect to the section $\sigma$ is given by
\begin{eqnarray}
\Sigma = \Omega^1_{ij} (dy^i\wedge dy^j) + \Omega^2_{i\mu} (dy^i\wedge dx^\mu) + \Omega^3_{\mu\nu} (dx^\mu\wedge dx^\nu), 
\end{eqnarray}
where $\Omega^1= s\chi+\frac{1}{2}[\chi, \chi]$ and $\Omega^2=s\phi +b\chi+[\phi,\chi]$. Here $[~,~]$ defines the Lie bracket. 
The Maurer--Cartan structural theorem states that the curvature $\Sigma$ is pure horizontal i.e. 
\begin{eqnarray}
\Omega^1=0,\qquad \Omega^2=0,
\end{eqnarray}
which provide the BRST transformations of fields in the theory. 
In this paper we will use this horizontality conditions to get the BRST and anti-BRST transformations 
of fields in  non-Abelian topologically massive $(B \wedge F)$ model.


\section{4D topologically massive  $(B \wedge F)$ theory} 

We first consider  the topologically massive Abelian model in $(3+1)$-dimensions of spacetime~\cite{crem,lah1,lah2} 
which contains a massive gauge field but keeping the gauge 
symmetry unbroken. In this model, the Abelian 1-form $A^{(1)} = dx^\mu A_\mu$ gauge field $A_\mu$ and antisymmetric  2-form
$B^{(2)} = \dfrac{1}{2!}\,(dx^\mu \wedge dx^\nu) B_{\mu\nu}$ field are coupled, in a physically meaningful manner, 
through a well-known topological $B\wedge F =\dfrac{1}{4}\,\varepsilon^{\mu\nu\eta\kappa} B_{\mu\nu} F_{\eta\kappa}$ 
term. Here $B_{\mu\nu}$ is the  Kalb-Ramond field and $F_{\mu\nu}=\partial_\mu A_\nu- \partial_\nu A_\mu$ is the field 
strength tensor corresponding to the  Abelian gauge field $A_\mu$. The mass of gauge field is put by hand in the model 
as a (constant) coupling parameter $m$ of the topological term. The  topologically massive Abelian model has the Lagrangian 
density\footnote{We adopt the conventions and notations such that the 4D flat Minkowski metric has 
mostly negative signatures: $\eta_{\mu\nu} = \eta^{\mu\nu} = \text{diga}\,(+1, -1, -1, -1)$. The Greek indices 
$\mu, \nu, \kappa,... = 0, 1, 2, 3$ correspond to spacetime directions whereas the Latin  indices $i, j, k,... = 1,2,3$  
stand for space directions only.}~\cite{crem,lah1,lah2}:
\begin{eqnarray}
{\cal L}_0 = -\frac{1}{4}\, F^{\mu\nu}F_{\mu\nu} + \frac{1}{12}\, H^{\mu\nu\kappa} H_{\mu\nu\kappa} 
+ \frac{m}{4}\,\varepsilon^{\mu\nu\eta\kappa}\, F_{\mu\nu}\, B_{\eta\kappa},
\label{lagrangian}
\end{eqnarray}
where $H_{\mu\nu\kappa}= \partial_\mu B_{\nu\kappa}+ \partial_\nu B_{\kappa\mu}+\partial_\kappa B_{\mu\nu}$ 
is the field strength of the Kalb-Ramond field.  The Abelian model is invariant under the following gauge transformations of the fields:
\begin{eqnarray}
A_\mu\to A_\mu+\partial_\mu \Omega , \qquad B_{\mu\nu}\to B_{\mu\nu},
\end{eqnarray}
and,
\begin{eqnarray}
B_{\mu\nu}\to B_{\mu\nu} - (\partial_\mu \Omega_\nu - \partial_\nu \Omega_\mu), \qquad A_\mu\to A_\mu,
\end{eqnarray}
where $\Omega(x)$ and $\Omega_\mu(x)$ are the local gauge transformation parameters which vanish at infinity. 
The Euler-Lagrange equations of motion for $A_\mu$ and $B_{\mu\nu}$ fields are give by, respectively 
\begin{eqnarray}
\partial_\mu F^{\mu\nu}&=&-\frac{m}{6}\, \varepsilon^{\nu\mu\eta\kappa}H_{\mu\eta\kappa},\nonumber\\
\partial_\mu H^{\mu\nu\eta}&= &\frac{m}{2}\,\varepsilon^{\nu\eta\kappa\rho} F_{\kappa\rho}.
\end{eqnarray}
After decoupling the above equations of motion for the fields, we get either 
\begin{eqnarray}
\left(\square + m^2\right) F_{\mu\nu}=0,
\label{kg1}
\end{eqnarray}
or, 
\begin{eqnarray}
\left(\square + m^2 \right) H^{\mu\nu\lambda}=0,
\label{kg2}
\end{eqnarray}
which are clearly the gauge-invariant  Klein-Gordon equations for massive $A_\mu$ and $B_{\mu\nu}$ fields.  
The counting of the degrees of  freedom shows that massive $B_{\mu\nu}$ field has 
three degrees of freedom as same as massive vector field $A_\mu$ in physical $(3+1)$-dimensions of spacetime. 


We now discuss the non-Abelian generalization of the above model. This theory is described by the 
following Lagrangian density\footnote{The dot and cross products in the $SU(N)$ algebraic space between two 
non-null vectors $X$ and $Y$ are defined as: $X \cdot Y = X^a Y^a,\; X \times Y = f^{abc} X^a Y^b T^c$. 
Here the structure constants $f^{abc}$ are chosen to be totally antisymmetric in their indices $a, b, c$ 
and $T^a$ are the 
generators of the gauge group $SU(N)$.}  \cite{mieg,oda,lah3,lah4} 
\begin{eqnarray}
{\cal L} = - \frac{1}{4}\, F_{\mu\nu} \cdot F^{\mu\nu} 
+ \frac{1}{12}\, H_{\mu \nu \eta}\cdot H^{\mu \nu \eta}
+ \frac{m}{4}\, \varepsilon^{\mu \nu \eta \kappa}\,B_{\mu \nu}\cdot F_{\eta\kappa},
\label{nonablag}
\end{eqnarray} 
where $F^a_{\mu \nu}\,T^a \equiv F_{\mu \nu} = \partial_\mu A_\nu 
- \partial_\nu A_\mu - g(A_\mu \times A_\nu)$ is the field strength 
tensor for the non-Abelian 1-form gauge field $A_\mu = A^a_\mu\,T^a$. The totally antisymmetric compensated curvature tensor 
$H_{\mu \nu \eta} \equiv H^a_{\mu\nu\eta} T^a$  for the 
non-Abelian gauge field $B_{\mu\nu} = B^a_{\mu \nu}\,T^a$ is defined as 
\begin{eqnarray} 
H^a_{\mu \nu \eta}\, T^a \equiv H_{\mu\nu\eta} &=&  D_\mu B_{\nu \eta} + D_\nu B_{\eta \mu} + D_\eta B_{\mu\nu} \nonumber\\
&+& g\,(F_{\mu \nu} \times K_\eta) + g\,(F_{\nu\eta} \times K_\mu) + g\,(F_{\eta \mu} \times K_\nu),
\label{cur2}
\end{eqnarray}
where 1-form $K^{(1)} = dx^\mu K_\mu \cdot T$  is the compensating auxiliary vector field  $K_\mu = K^a_\mu  T^a$ and $g$ is a dimensionless coupling 
constant. The gauge bosons $A_\mu$ acquire mass through
the topological  term  $(B \wedge F)$ without taking any help of Higgs mechanism. 
The presence of topological term
$\frac{m}{4}\,\varepsilon^{\mu\nu\rho\sigma}B_{\mu\nu} \cdot F_{\rho\sigma}$ also
ensures us the  CP-invariance of the model. It is because of the parity transformation of Kalb-Ramond field:
$B_{0i}\to - B_{0i}$, $B_{ij}\to B_{ij}$ ~\cite{kr:1973}.
The topological term does not break Lorentz invariance in $(3+1)$-dimensions unlike the topological term present in \cite{cbj:1990}. 
The compensating auxiliary vector field is required for the invariance of  kinetic term for tensor field $B_{\mu\nu}$ 
under the non-Abelian vector gauge transformation:
$B_{\mu\nu}\to B_{\mu\nu}- (D_\mu \Lambda_\nu - D_\nu \Lambda_\mu)$ (see below). The absence of  propagator of the auxiliary vector
field in Eq. (\ref{nonablag}) implies the absence of its role in the physical processes. We will see from the BRST transformation of 
$K_\mu$ that its all  modes are unphysical. This model is shown to be renormalizable algebraically in~\cite{lah4} and unitary at tree level.

The non-Abelian generalization  must keep all the symmetries that were present in the Abelian model. 
The above Lagrangian density respects two kinds of gauge symmetry transformations: (i) scalar gauge symmetry ($\delta_1$), and 
(ii) vector gauge symmetry $(\delta_2)$. These symmetry transformations are listed as follows:
\begin{eqnarray}
&&\delta_1 A_\mu = D_\mu \zeta \equiv \partial_\mu \zeta - g(A_\mu \times \zeta),\nonumber \label{vecgaugeA}\\
&&\delta_1 B_{\mu\nu} = - g(B_{\mu\nu} \times \zeta), \qquad \delta_1 K_\mu = - g(K_\mu \times \zeta), \nonumber\\
&& \delta_1 F_{\mu\nu} = -g (F_{\mu\nu} \times \zeta), \qquad \delta_1 H_{\mu\nu\eta} = - g(H_{\mu\nu\eta} \times \zeta), \nonumber\\
&&\nonumber\\
&& \delta_2 B_{\mu\nu} = -(D_\mu \Lambda_\nu - D_\nu \Lambda_\mu ), \quad  \qquad \delta_2 K_\mu = - \Lambda_\mu,  \nonumber\\
&& \delta_2 A_\mu = 0, \qquad \delta_2 F_{\mu \nu}=0, \qquad \delta_2 H_{\mu \nu \eta}=0.
\end{eqnarray}
where $\zeta = \zeta \cdot T$ and $\Lambda_\mu = \Lambda_\mu \cdot T$  are the
$SU(N)$-valued  local ``scalar''  and ``vector" gauge transformation parameters. 
Under these local gauge transformations, the Lagrangian density transforms as
\begin{eqnarray}
\delta_1 {\cal L} = 0, \qquad \delta_2 {\cal L} = -  \partial_\mu\, \Bigl [\frac{m}{2}\,\varepsilon^{\mu\nu\eta\kappa} 
\Lambda_\nu \cdot F_{\eta\kappa}\Bigr].
\end{eqnarray}
Thus, the action integral ($S = \int d^4x {\cal L}$) remains invariant under the gauge transformations for the 
physically well-defined fields which vanish rapidly at infinity due to Gauss divergence theorem. Also, the combined 
gauge transformations $\delta = (\delta_1 + \delta_2)$ leaves the action integral invariant.


\section{Off-shell nilpotent and absolutely anticommuting (anti-)BRST symmetries: geometrical superfield formalism}

In this section, we derive the complete set of off-shell nilpotent and absolutely anticommuting (anti-)BRST 
symmetries with the help of Bonora--Tonin superfield approach to BRST formalism. For this purpose, we  generalize our
ordinary 4D spacetime to the $(4, 2)$D superspace.  The latter is characterized   by a pair of Grassmannian 
variables\footnote{The Grassmannian variables obey
the following Hermiticity properties: $\theta^\dagger = \theta$ and ${\bar \theta}^\dagger = - \bar\theta$.} 
($\theta, \bar \theta$) (with $\theta^2 = \bar \theta^2 = 0,\; \theta \bar \theta + \bar \theta \theta = 0$) 
in addition to the bosonic spacetime variables $x^\mu$ (with $\mu = 0,1,2,3$) as~\cite{bt1,bt2,del}
\begin{eqnarray}
x^\mu \to Z^M \equiv (x^\mu, \theta, \bar \theta), 
\qquad \partial_\mu \to \partial_M \equiv (\partial_\mu, \partial_\theta, \partial_{\bar \theta}),
\end{eqnarray} 
where the super-coordinates $Z^M$ parametrized the $(4, 2)$D supermanifold. The partial  derivatives $\partial_\theta = \frac{\partial}{\partial \theta}$
and $\partial_{\bar \theta} = \frac{\partial}{\partial \bar \theta}$ 
(with $\partial^2_\theta = \partial^2_{\bar \theta} = 0,\; \partial_\theta \partial_{\bar \theta} + \partial_{\bar \theta} \partial_\theta =0$) 
are the translational generators along the Grassmannian directions $\theta$ and $\bar \theta$, respectively. We shall see later on that
these translational generators provide the geometrical meaning of the anti-BRST and BRST symmetry transformations, respectively.

In our upcoming subsections, we shall exploit the horizontality conditions and integrability condition for the 
derivation of proper (anti-)BRST transformations.

\subsection{Derivation of the (anti-)BRST transformations of YM field and corresponding ghost fields}

For the derivation of (anti-)BRST symmetry transformations of the YM gauge field,  we generalize the exterior derivative $d = dx^\mu\partial_\mu$ 
(with $d^2 = 0$) and 1-form connection $A^{(1)} = dx^\mu A^a_\mu T^a$ to the super-exterior derivative $\tilde d$ (with $\tilde d^2 = 0$) and 
super 1-form $\tilde {\cal A}^{(1)} = dx^\mu \tilde {\cal A}^a_\mu T^a$ on the $(4, 2)$D supermanifold in the following fashion:         
\begin{eqnarray}
\tilde d = d Z^M \partial_M &\equiv& dx^\mu \,\partial_\mu + d \theta \,\partial_\theta
+ d \bar \theta \,\partial_{\bar\theta}, \nonumber\\
\tilde {\cal A}^{(1)} = d Z^M A_M &\equiv& dx^\mu \,\tilde {\cal A}_\mu (x,\theta,\bar\theta) 
+ d \theta \,  {\tilde {\bar{\cal F}}} (x,\theta,\bar\theta) + {d \bar\theta}\, {\tilde {\cal F}} (x,\theta,\bar\theta ), 
\end{eqnarray} 
where the superfields $\tilde {\cal A}_\mu (x,\theta,\bar\theta)$, $\tilde {\cal F} (x,\theta,\bar\theta)$
and ${\tilde {\bar {\cal F}}}(x,\theta,\bar\theta)$, as the super-multiplets of super 1-form, are the generalization of 1-form gauge field
$A_\mu (x)$, ghost field $C (x)$ and anti-ghost field $\bar C (x)$, respectively, 
on the $(4, 2)$D supermanifold. One can expand these superfields along the Grassmannian directions $(\theta, \bar \theta)$ as
\begin{eqnarray}
\tilde {\cal A}_\mu (x, \theta, \bar \theta) &=& A_\mu (x) + \theta \bar R_\mu (x) + \bar \theta R_\mu (x)
+ \theta \bar\theta S_\mu (x), \nonumber\\
\tilde {\cal F} (x, \theta, \bar \theta) &=& C (x) + \theta  \bar B_1 (x) + \bar\theta B_1 (x) + \theta \bar\theta s(x), \nonumber\\
{\tilde {\bar {\cal F}}} (x, \theta, \bar \theta) &=& \bar C (x) + \theta \bar B_2 (x) +  \bar\theta B_2 (x)
+ \theta \bar\theta \bar s (x),
\label{super-ex1}
\end{eqnarray}
where the secondary fields $\bar R_\mu, R_\mu, s, \bar s$ are fermionic and the
remaining  secondary fields $S_\mu, B_1, \bar B_1, B_2, \bar B_2$ are bosonic in nature.

To determine the values of these secondary fields, we invoke the following HC 
\begin{eqnarray}
&& \tilde d \tilde {\cal A}^{(1)} + \frac{i}{2}\,g  \left[\tilde {\cal A}^{(1)},\, \tilde{\cal A}^{(1)} \right]
= d A^{(1)} + \frac{i}{2} \,g \left[A^{(1)},\,  A^{(1)}\right] \Longrightarrow \tilde {\cal F}^{(2)} = F^{(2)},  
\label{hc1}
\end{eqnarray}
where the super 2-form $\tilde {\cal F}^{(2)} = \frac{1}{2!} (dZ^M \wedge dZ^N)\; \tilde{\cal F}_{MN}$ is the generalization of 
$F^{(2)} = \frac{1}{2!} (dx^\mu \wedge dx^\nu) F_{\mu\nu}$ on the supermanifold. The HC in the literature is also known as 
{\it soul-flatness} condition which states that the r.h.s. is independent of the Grassmannian variables when it is generalized onto $(4,2)$D supermanifold.
To be more precise, the HC demands that all the Grassmannian  components of the super 2-form curvature $\tilde{\cal F}_{MN}$ are equal to 
zero (i.e. $\tilde{\cal F}_{\mu\theta}$ = $\tilde{\cal F}_{\mu \bar\theta}$ = $\tilde{\cal F}_{\theta\theta}$ 
= $\tilde{\cal F}_{\bar \theta \bar \theta}$ = $\tilde{\cal F}_{\theta \bar\theta} = 0$).
As we already know that the kinetic term ($-\frac{1}{4}\, F_{\mu\nu} \cdot F^{\mu\nu}$) 
for the gauge field $A_\mu$ remains invariant under the combined gauge transformations $(\delta)$. Thus, the kinetic term would also remain 
invariant under (anti-)BRST transformations. Physically, the HC implies that the gauge-invariant quantity must be independent 
of the Grassmannian variables $(\theta, \bar \theta)$ (i.e. $-\frac{1}{4}\, \tilde {\cal F}_{MN} \cdot \tilde {\cal F}^{MN} = -\frac{1}{4}\, F_{\mu\nu} \cdot F^{\mu\nu}$) when it is generalized on the $(4,2)$ supermanifold. It is worthwhile to point out that the Grassmannian variables 
are just a mathematical artifact and they cannot be physically realized in our physical $4$D spacetime.  
In fact, they are used to construct the $(4, 2)$-dimensional superspace.

By exploiting the above HC, we obtain the values of the secondary fields  [cf. (\ref{sec1})].    
The substitution of the values of secondary fields in the super-expansions of the superfields, 
we obtain\footnote{The Nakanishi-Lautrup fields $B, \bar B$ are real and the anticommuting ghost fields satisfy the following Hermiticity properties:
$C^\dagger = C$, and ${\bar C}^\dagger  = - \bar C.$}
\begin{eqnarray}
\tilde {\cal A}^{(h)}_\mu (x, \theta, \bar \theta) &= & A_\mu + \theta D_\mu \bar C  + \bar\theta D_\mu C 
+ \theta \bar\theta \left(D_\mu B -  g (D_\mu C \times \bar C)\right),\nonumber\\
\tilde {\cal F}^{(h)} (x, \theta, \bar \theta) &=& C + \theta  \bar B + \bar\theta \,\frac{g}{2} \big(C \times C \big) 
- \theta \bar\theta g \big(\bar B \times C \big), \nonumber\\
{\tilde {\bar {\cal F}}}^{(h)} (x, \theta, \bar \theta) & = & \bar C  + \theta \,\frac{g}{2} \big(\bar C \times \bar C \big) 
 + \bar\theta \,B + \theta \bar\theta g \big(B \times \bar C\big),
\label{sup1}
\end{eqnarray}
where the superscript $(h)$ on the superfields denotes the super-expansions obtained after the application of HC (\ref{hc1}).  
We have  made the identifications: $ \bar B_1 = \bar B$ and  $B_2 = B$ for the Nakanishi--Lautrup (NL) fields $B$ and $\bar B$. These fields 
are required for the off-shell nilpotency of the (anti-)BRST transformations.  
Similarly, the super-curvature $\tilde {\cal F}^{(h)}_{\mu\nu}$ corresponding to the superfield ${\cal A}^{(h)}_\mu$ can be written as 
\begin{eqnarray}
\tilde {\cal F}^{(h)}_{\mu\nu} (x,\theta,\bar\theta) = F_{\mu\nu} &-& \theta g\big(F_{\mu\nu} \times \bar C\big)
- \bar \theta g \big(F_{\mu\nu} \times  C \big) \nonumber\\
&+& \theta\,\bar \theta \big(g^2 (F_{\mu\nu} \times C) \times \bar C 
- g (F_{\mu\nu} \times B)\big).
\label{sup2}
\end{eqnarray}
From the above super-expansions, one can easily read-off all
the (anti-)BRST transformations for the YM field and corresponding (anti-)ghost fields. These are listed as follows 
\begin{eqnarray}
&& s_b A_\mu = D_\mu C, \quad s_b C = \frac{g}{2} (C \times C), \quad s_b \bar C = B,   \quad s_b B = 0, \nonumber\\
&& s_b \bar B = -g (\bar B \times C),  \quad  s_b F_{\mu\nu} = - g(F_{\mu\nu} \times C), \nonumber\\
&& \nonumber\\
&& s_{ab} A_\mu = D_\mu \bar C, \quad s_{ab} \bar C = \frac{g}{2} (\bar C \times \bar C), \quad s_{ab} C = \bar B, \quad s_{ab} \bar B = 0, \nonumber\\
&&  s_{ab} B = - g(B \times \bar C), \quad s_{ab} F_{\mu\nu} = - g(F_{\mu\nu} \times \bar C). 
\label{sab1}
\end{eqnarray}
Geometrically, the BRST transformation ($s_b$) for any generic field $\Sigma(x)$ is equivalent to the translational of corresponding 
superfield $\tilde \Sigma^{(h)}(x, \theta, \bar \theta)$ along $\bar \theta$-direction while keeping $\theta$-direction fixed. 
In a similar fashion, the anti-BRST transformation $(s_{ab})$ can be obtained  by taking the translational of the 
superfield along  $\theta$-direction while  $\bar \theta$-direction  remains intact.  As a consequence, the following 
mappings are valid between the Grassmannian translational generators ($\partial_\theta, \partial_{\bar\theta}$)
 and the (anti-)BRST symmetry transformations, namely;
\begin{eqnarray}
&& \frac{\partial}{\partial \bar\theta}\, \tilde \Sigma^{(h)} (x, \theta, \bar\theta)\Big|_{\theta = 0} = s_b \,\Sigma (x), 
\qquad  \frac{\partial}{\partial \theta}\, \tilde \Sigma^{(h)} (x, \theta, \bar\theta) \Big |_{\bar \theta = 0} = s_{ab} \, \Sigma (x), \nonumber\\
&& \frac{\partial}{\partial \bar\theta}\, \frac{\partial}{\partial \theta}\, \tilde \Sigma^{(h)} (x, \theta, \bar\theta)
= s_b s_{ab} \,\Sigma (x).
\end{eqnarray}
The (anti-)BRST transformations of the NL auxiliary fields $B$ and $\bar B$ have been derived from the requirements of 
the nilpotency and absolute anticommutativity of the (anti-)BRST transformations.

We point out that the absolute anticommutativity property of the BRST and anti-BRST transformations is satisfied due to the validity of the following
CF condition~\cite{cur} [cf. (\ref{sec1})]:
\begin{eqnarray}
B + \bar B - g \big(C \times \bar C\big) = 0. \label{cf1}
\end{eqnarray}
It is a physical condition on the theory in the sense that it is BRST as well as anti-BRST invariant quantity 
(i.e. $s_{(a)b} [B + \bar B - g \big(C \times \bar C\big)] = 0$). This is an original CF condition which was emerged automatically first time
for the non-Abelian 1-form gauge theory within the 
framework of superfield approach to BRST formalism \cite{bt1}. For the sake of brevity, the restriction  $\tilde {\cal F}_{\theta \bar \theta} = 0$ leads to
the above CF condition.

\subsection{(Anti-)BRST symmetries of antisymmetric gauge field and associated ghost fields}

In this subsection, we focus on the derivation of the BRST and  anrt-BRST transformations for $B_{\mu\nu}$ 
and corresponding (anti-)ghost fields. For this purpose, we use another HC as given below
\begin{eqnarray}
H^{(3)} = \tilde {\cal H}^{(3)}, \label{hc2}
\end{eqnarray} 
which again implies that the kinetic term for the 2-form field $B_{\mu\nu}$ is a gauge-invariant quantity.
Here  $\tilde {\cal H}^{(3)} = \frac{1}{3!}\,(dZ^L \wedge dZ^M \wedge dZ^N){\cal H}_{LMN}$ defines the 3-form super-curvature on the $(4, 2)$D supermanifold 
corresponding to the 3-form curvature $H^{(3)} = \frac{1}{3!}\,(dx^\mu \wedge dx^\nu \wedge dx^\eta)H_{\mu\nu\eta}$. These (super-)curvature are
defined in the following fashion:
\begin{eqnarray}
\tilde {\cal H}^{(3)} &=& \tilde {d} \tilde {\cal B}^{(2)}+ ig \left[\tilde {\cal A}^{(1)}_{(h)}, \, \tilde {\cal B}^{(2)} \right] 
+i g \left[\tilde {\cal K}^{(1)}, \, \tilde {\cal F}^{(2)}_{(h)} \right],\nonumber\\
H^{(3)} &=& dB^{(2)} + ig \left[A^{(1)}, \, B^{(2)} \right] + ig \left[K^{(1)}, \, F^{(2)} \right],  
\end{eqnarray}
where $\tilde {\cal A}^{(1)}_{(h)}$ is the super 1-form obtained after the application of first HC (\ref{hc1}) and $\tilde {\cal F}^{(2)}_{(h)}$ defines the corresponding super-curvature.
It is straightforward to check that $H^{(3)}$ produces the curvature tensor (\ref{cur2}). The super 1-form
$\tilde {\cal K}^{(1)}$ and super 2-form $\tilde {\cal B}^{(2)}$ can be written
as follows: 
\begin{eqnarray}
\tilde {\cal K}^{(1)}&=& dx^{\mu}\,\tilde {\cal K}_{\mu}(x, \theta, \bar\theta) 
+ d\theta \,{\tilde {\bar \xi}}(x, \theta, \bar\theta) + {d\bar\theta} \,\tilde \xi(x,\theta,\bar\theta), \nonumber\\
\tilde {\cal B}^{(2)}&=&\frac{1}{2!}\, (dZ^M \wedge dZ^N)\,\tilde {\cal B}_{MN}(x, \theta,\bar\theta)\nonumber\\
&\equiv & \frac{1}{2!} \,(dx^{\mu}\wedge dx^{\nu})\, \tilde {\cal B}_{\mu \nu}(x, \theta,\bar\theta) 
+(dx^{\mu} \wedge d\theta)\, {\tilde {\bar{\cal F}}}_{\mu}(x,\theta,\bar\theta)
+(dx^\mu \wedge d\bar\theta) \,\tilde {\cal F}_\mu (x,\theta,\bar\theta)\nonumber\\
&+&(d\theta \wedge d\bar\theta)\,\tilde\Phi(x,\theta,\bar\theta)
+(d\theta \wedge d\theta)\,\tilde{\bar \beta}(x,\theta,\bar\theta) 
+ (d\bar\theta \wedge d\bar\theta)\,\tilde \beta(x,\theta,\bar\theta),
\end{eqnarray}
where $\tilde {\cal K}^{(1)}$ and $\tilde {\cal B}^{(2)}$ are the generalizations of $K^{(1)}$ and $ B^{(2)}$, respectively on the supermanifold.
Again, the super-multiples, as the components of the above super 1-form and super 2-form, can be expanded along the directions of Grassmannian variables
$(\theta, \bar \theta)$ as 
\begin{eqnarray}
\tilde {\cal B}_{\mu\nu} (x, \theta, \bar \theta) &=& B_{\mu\nu} (x) + \theta  \bar R_{\mu\nu} (x) + \bar \theta R_{\mu\nu} (x)
+ \theta \bar\theta  S_{\mu\nu} (x), \nonumber\\
\tilde {\cal K}_\mu (x, \theta, \bar \theta) &=& K_\mu (x) + \theta \; \bar P_\mu (x) + \bar\theta \, 
P_\mu (x) + \theta \bar\theta\,Q_\mu (x), \nonumber\\
\tilde {\cal F}_{\mu}(x, \theta, \bar\theta) &=& C_\mu (x)+ \theta \,{\bar b}^{(1)}_\mu(x) 
+ \bar \theta \, b^{(1)}_\mu(x) + \theta \bar\theta \,q_\mu(x), \nonumber\\
{\tilde {\bar{\cal F}}}_{\mu}(x, \theta, \bar\theta) &=& {\bar C}_\mu (x)+ \theta \, \bar b^{(2)}_\mu(x) 
+ \bar \theta \, b^{(2)}_\mu(x) + \theta \bar\theta \, {\bar q}_\mu(x),  \nonumber\\
\tilde \Phi(x, \theta, \bar\theta)&=& \phi(x) + \theta \, {\bar f}_1(x) + \bar\theta \, f_1(x) 
+ \theta \bar\theta \, b_1(x), \nonumber \\
\tilde \beta (x, \theta, \bar\theta)&=& \beta(x) + \theta \,{\bar f}_2(x) + \bar\theta \, f_2(x) 
+  \theta \bar\theta \, b_2(x), \nonumber\\
\tilde {\bar \beta} (x, \theta, \bar\theta)&=& {\bar \beta}(x) + \theta \, {\bar f}_3(x) + \bar\theta \, f_3(x) 
+ \theta \bar\theta \, b_3(x), \nonumber\\
\tilde {\xi} (x, \theta, \bar\theta)&=& \xi(x) + \theta \ {\bar R}_1(x) + \bar\theta \, R_1(x) 
+  \theta \bar\theta \, S_1(x), \nonumber\\
\tilde {\bar{\xi}} (x, \theta, \bar\theta)&=& {\bar \xi}(x) + \theta \, {\bar R}_2(x) + \bar\theta \, R_2(x) 
+  \theta \bar\theta \,{S}_2(x),
\label{sup3}
\end{eqnarray}
where the secondary fields $R_{\mu\nu}$, $\bar R_{\mu\nu}$, $P_\mu$, $\bar P_\mu$, $q_\mu$, $\bar q_\mu$, $f_1$, $\bar f_1$, 
$f_2$, $\bar f_2$, $f_3$, $\bar f_3$, $S_1$, $S_2$ are fermionic in nature and $S_{\mu\nu}$, $Q_\mu$, $b^{(1)}_\mu$, $\bar b^{(1)}_\mu$, 
$b^{(2)}_\mu$, $\bar b^{(2)}_\mu$, $b_1$, $b_2$, $b_3$, $R_1$, $\bar R_1$, $R_2$, $\bar R_2$ are the  bosonic secondary fields.

By using the second HC (\ref{hc2}) together with (\ref{sup1}) and (\ref{sup2}), we obtain the values of the above secondary fields 
except $P_\mu$, $\bar P_\mu$ and $Q_\mu$ [cf. (\ref{sec2})]. As a result, we get the desired super-expressions of the above 
superfields (\ref{sup3}):   
\begin{eqnarray*}
\tilde {\cal B}_{\mu\nu}^{(h)}(x,\theta,{\bar\theta}) &=& 
B_{\mu\nu} + \theta \, \bigl[ - (D_\mu \bar C_\nu - D_\nu \bar C_\mu) + g(\bar C \times B_{\mu\nu}) + g(\bar \xi \times F_{\mu\nu}) \big] \nonumber\\ 
&+& {\bar\theta} \,  \bigl[- (D_\mu C_\nu -D_\nu C_\mu) + g(C \times B_{\mu\nu}) + g(\xi \times F_{\mu\nu}) \bigr] \nonumber\\ 
&+& \theta \,\bar\theta \, \bigl [- (D_\mu B_\nu - D_\nu B_\mu) + g(D_\mu C \times  \bar C_\nu) 
- g(D_\nu C \times \bar C_\mu)  \nonumber\\
&+& g(B \times B_{\mu\nu}) + g^2 \big(\bar \xi \times (F_{\mu\nu} \times C)\big) - g^2 \big(\bar C\times (C \times B_{\mu\nu}) \big) \nonumber\\
&+& g \big(\bar C \times (D_\mu C_\nu - D_\nu C_\mu) \big) 
+ g(R \times F_{\mu\nu}) - g^2\big(\bar C \times (\xi \times F_{\mu\nu})\big) \bigr],\nonumber\\ 
\tilde {\cal F}_\mu^{(h)}(x,\theta,{\bar\theta}) &=& C_\mu + \theta \,{\bar B}_\mu  
+ {\bar\theta} \, \big[- D_\mu \beta + g(C \times C_\mu)\big] \nonumber\\
&+& \theta \, \bar \theta \bigl[D_\mu \lambda -g (D_\mu \bar C \times  \beta) 
- g(\bar B \times C_\mu) - g(\bar B_\mu \times C) \big],  \nonumber\\ 
{\tilde {\bar {\cal F}}}_\mu^{(h)}(x,\theta,{\bar\theta}) &=& {\bar C}_\mu
+ \theta \, \big[- D_\mu \bar \beta + (\bar C \times \bar C_\mu)\big] + \bar\theta \,B_\mu \nonumber\\
&+& \theta\,\bar\theta \, \big[- D_\mu \bar \lambda + g(D_\mu C \times \bar \beta) 
+ g(B \times \bar C_\mu) + g(B_\mu \times \bar C)\big], \nonumber\\ 
\end{eqnarray*}
\begin{eqnarray}
\tilde \beta^{(h)}(x,\theta,{\bar\theta}) &=& \beta + \theta \,\lambda
+ \bar \theta\, g\big(C \times \beta \big) 
+ \theta\, \bar \theta \,\big[g(C \times \lambda) - g(\bar B \times \beta)\big], \nonumber\\
\tilde {\bar \beta}^{(h)}(x,\theta,\bar\theta ) &=& \bar \beta + \theta \, g\big(\bar C \times \bar \beta \big)
+ \bar \theta \, \bar \lambda 
+ \theta \, \bar \theta \, \big[-g(\bar C \times \bar \lambda) + g(B \times \bar \beta)\big], \nonumber\\
\tilde \Phi^{(h)}(x,\theta,\bar\theta ) &=& \phi + \theta \bar \rho
 + \bar\theta \rho  \nonumber\\
&+& \theta\, \bar \theta  \big[g(B \times \phi) - g(\bar C \times \bar \rho)
- g(C \times \bar \lambda) + g^2 \big(C \times (C \times \bar \beta)\big) \big],\nonumber\\  
\tilde {\xi}^{(h)}(x,\theta,\bar\theta ) &=& \xi + \theta \, \bar R
+ \bar\theta\,  \big[- \beta + (C \times \xi)\big] 
+ \theta \,\bar\theta \, \big[\lambda - g(\bar R \times C) - g(\bar B \times \xi)\big],  \nonumber\\
\tilde {\bar{\xi}}^{(h)}(x,\theta,\bar\theta ) &=& \bar \xi + \theta \, \big[- \bar \beta + (\bar C \times \bar \xi)\big]
+ \bar\theta \, R + \theta \,\bar\theta\, \big[-\bar \lambda - g(R \times \bar C) - g(B \times \bar \xi)\big]. \qquad
\label{sup4}
\end{eqnarray}
In the above, we have chosen $b^{(2)}_\mu = B_\mu$, $\bar b^{(1)}_\mu = \bar B_\mu$, $R_2 = R$, $\bar R_1 = \bar R$ for the bosonic  
NL-type auxiliary fields $B_\mu$, $\bar B_\mu$, $R$, $\bar R$ and $f_1 = \rho$, $\bar f_1 = \bar \rho$, $\bar f_2 = \lambda$, $f_3 = \bar \lambda$ 
for the additional fermionic NL-type fields $\rho$, $\bar \rho$, $\lambda$, $\bar \lambda$.  
Again, these (bosonic) fermionic auxiliary fields are required for the off-shell nilpotency of the (anti-)BRST transformations.
One can also express the 3-form super-curvature in terms of the Grassmannian variables as
\begin{eqnarray}
\tilde {\cal H}^{(h)}_{\mu\nu\eta}(x, \theta, \bar\theta) &=& H_{\mu\nu\eta} 
- \theta g (H_{\mu\nu\eta} \times \bar C) - \bar \theta  g(H_{\mu\nu\eta} \times C) \nonumber\\ 
&+& \theta \bar \theta \big[- g(H_{\mu\nu\eta} \times B) + g^2(H_{\mu\nu\eta} \times C) \times \bar C \big]. 
\label{sup5}
\end{eqnarray}
As a consequence of the above super-expansions, we obtain the following BRST and anti-BRST symmetry transformations, namely;
\begin{eqnarray*}
&& s_b B_{\mu\nu} = - (D_\mu C_\nu -D_\nu C_\mu) + g(C \times B_{\mu\nu}) + g(\xi \times F_{\mu\nu}),\quad
s_b C_{\mu} = - D_\mu \beta + g(C \times C_\mu), \nonumber\\
&& s_b \bar C_\mu = B_\mu,  \quad  s_b \beta = g(C \times \beta), \quad s_b \bar \beta = \bar \lambda, 
\quad  s_b \phi = \rho, \quad s_b \xi = - \beta + g(C \times \xi), \nonumber\\
&& s_b \bar \xi = R,  \quad s_b \bar R = \lambda - g(\bar R \times C) - g(\bar B \times \xi), 
\quad s_b \lambda = g(\lambda \times C) -  g(\bar B \times \beta),\nonumber\\
&& s_b \bar \rho = g(B \times \phi) +  g(\bar \rho \times C)- g(\rho \times \bar C) 
- g^2 \big(C \times (\bar C \times \phi) \big), \nonumber\\
&&s_b \bar B_\mu = - D_\mu \big(\rho - g (C \times \phi) \big) + g \big(B - g (C \times \bar C) \big) \times C_\mu - g (\bar B_\mu \times C) 
+ g(\bar C \times D_\mu \beta) \nonumber\\
&& s_b H_{\mu\nu\eta} = - g (H_{\mu\nu\eta} \times C), \qquad s_b[B_\mu, \, R, \, \rho, \,\bar \lambda] = 0,
\end{eqnarray*}
\begin{eqnarray}
&& s_{ab} B_{\mu\nu} = - (D_\mu \bar C_\nu -D_\nu \bar C_\mu) + g(\bar C \times B_{\mu\nu}) + g(\bar \xi \times F_{\mu\nu}),\;\;
s_{ab} \bar C_{\mu} = - D_\mu \bar\beta + g(\bar C \times \bar C_\mu), \nonumber\\
&& s_{ab} C_\mu = \bar B_\mu,  \quad  s_{ab} \bar \beta = g(\bar C \times \bar \beta), \quad s_{ab}  \beta = \lambda, 
\quad  s_{ab} \phi = \bar \rho, \quad s_{ab} \bar \xi = - \bar \beta + g(\bar C \times \bar \xi), \nonumber\\
&& s_{ab} \xi = \bar R,  \quad s_{ab} R = \bar \lambda - g(R \times \bar C) - g(B \times \bar \xi), 
\quad s_{ab} \bar \lambda = g(\bar \lambda \times \bar C) -  g(B \times \bar \beta), \nonumber\\
&& s_{ab} \rho = g(\bar B \times \phi) -  g(\bar \rho \times C)+ g(\rho \times \bar C) 
- g^2 \big(\bar C \times (C \times \phi) \big), \nonumber\\
&& s_{ab} B_\mu = - D_\mu \big(\bar \rho - g(\bar C \times \phi)\big) + g \big(\bar B -g (C \times \bar C) \big)\times \bar C_\mu - g (B_\mu \times \bar C) 
+ g(C \times D_\mu \bar \beta) \nonumber\\
&& s_{ab} H_{\mu\nu\eta} = - g (H_{\mu\nu\eta} \times \bar C), \qquad s_{ab}[\bar B_\mu, \, \bar R, \, \bar \rho, \,\lambda] = 0.
\label{sab2}
\end{eqnarray}
These transformations are also off-shell nilpotent and absolutely anticommuting in nature.


\subsection{(Anti-)BRST transformations of $K_\mu$ and associated ghosts}

We have, so far,  determined the BRST and anti-BRST transformations for the YM, Kalb-Ramond and their associated (anti-)ghost fields. But the proper 
(anti-)BRST transformations of the compensating auxiliary vector field are still unknown. This is because of the fact that   
the second HC is incapable to determine the precise value of the secondary fields $P_\mu$, $\bar P_\mu$ and $Q_\mu$.

It is to be noted that the field strength tensors transform covariantly (i.e.  $\delta F_{\mu\nu} = - g(F_{\mu\nu} \times \zeta )$ and 
$\delta H_{\mu\nu\eta} = - g (H_{\mu\nu\eta} \times \zeta)$) under the 
combined gauge transformations $\delta$. In a similar manner, it is interesting to point out that the following quantity
\begin{eqnarray}
\delta[(D_\mu K_\nu - D_\nu K_\mu) - B_{\mu\nu}] = - [(D_\mu K_\nu - D_\nu K_\mu) - B_{\mu\nu}] \times \zeta,
\end{eqnarray}
transforms covariantly under the combined gauge transformations, too.

In the language of differential forms, one can write 
\begin{eqnarray}
&& d K^{(1)} + ig \left[A^{(1)},\, K^{(1)}) \right] - B^{(2)} = \frac{1}{2!}\, (dx^\mu \wedge dx^\nu)
[(D_\mu K_\nu - D_\nu K_\mu) - B_{\mu\nu}],
\end{eqnarray}
which is clearly a 2-form quantity. Generalizing this 2-form quantity on the $(4,2)$D superspace which in turn produces the third HC 
\begin{eqnarray}
&& \tilde d \tilde {\cal K}^{(1)} + ig \left[\tilde {\cal A}^{(1)}_{(h)},\, \tilde {\cal K}^{(1)} \right]  - \tilde {\cal B}^{(2)}_{(h)} 
= d K^{(1)} + ig \left[A^{(1)},\,  K^{(1)} \right] - B^{(2)}.
\label{hc3}
\end{eqnarray}
It is worthwhile to mention that the above HC can also be obtained  from the integrability of the second HC (\ref{hc2}) \cite{mieg}.
Exploiting the above HC and setting all the Grassmannian differential equal to zero, we obtain the precise values of the renaming secondary fields 
[cf. (\ref{sec3})] and we have the following super-expansion of $\tilde {\cal K}_\mu$ as given below 
\begin{eqnarray}
\tilde {\cal K}_\mu^{(h)}(x,\theta,\bar\theta ) &=& K_\mu + \theta \, 
\big[D_\mu \bar \xi  - \bar C_\mu - g(K_\mu \times \bar C)\big] 
+ \bar\theta \, \big[D_\mu \xi - C_\mu - g(K_\mu \times C)\big]  \nonumber\\ 
&+&\theta\,\bar\theta \, \big[D_\mu R - B_\mu 
- g(D_\mu C \times \bar \xi) - g (K_\mu \times B) \nonumber\\
&-& g\big(D_\mu \xi - C_\mu -g( K_\mu \times C)\big)\times \bar C \big].
\end{eqnarray}
Thus, we obtain the following BRST and anti-BRST transformations for the compensating auxiliary field: 
\begin{eqnarray}
&& s_b K_\mu = D_\mu \xi -  C_\mu - g(K_\mu \times C), \qquad s_{ab} K_\mu = D_\mu \bar \xi - \bar C_\mu - g(K_\mu \times \bar C).
\label{sab3}
\end{eqnarray}
The above transformations as listed in (\ref{sab2}) and (\ref{sab3}) are off-shell nilpotent and absolutely anticommuting. 
However, the absolute anticommutativity property is satisfied on the constrained hypersurface defined by the CF-type condition 
(\ref{cf1}) and the following additional CF-type conditions [cf. (\ref{anticom2})]:      
\begin{eqnarray}
&& \bar B_\mu + B_\mu + D_\mu \phi - g(\bar C \times C_\mu) - g(C \times \bar C_\mu) = 0, \nonumber\\
&& \bar R + R + \phi - g(\bar C \times \xi) - g(C \times \bar \xi) = 0, \nonumber\\
&& \rho + \lambda - g(C \times \phi) - g(\bar C \times \beta) = 0, \nonumber\\
&& \bar \rho + \bar \lambda - g(\bar C \times \phi) - g(C \times \bar \beta) = 0.
\label{cf2}
\end{eqnarray}
These CF-type conditions emerge from the second and third HCs [cf. (\ref{sec2}) and (\ref{sec3})]. 
Furthermore, it is to be noted that the first two CF-type conditions are bosonic whereas last two are fermionic in nature.

\section{Coupled but equivalent Lagrangian densities}
Using the basic principles and ingredients of BRST formalism, the most appropriate (anti-)BRST invariant Lagrangian densities which incorporate the 
gauge-fixing and Faddeev--Popov ghosts terms can be written as
\begin{eqnarray}
{\cal L}_{(B)} = {\cal L}
+ s_b s_{ab} \bigg[\frac{1}{2}\, A_\mu \cdot A^\mu + \bar C \cdot C + \frac{1}{2}\, \phi \cdot \phi 
+2\, \bar\beta \cdot \beta + \bar C_\mu \cdot C^\mu - \frac{1}{4}\, B^{\mu\nu} \cdot B_{\mu\nu}\bigg], \quad
\label{LB}
\end{eqnarray}
\begin{eqnarray}
{\cal L}_{(\bar B)} = {\cal L}
- s_{ab} s_b \bigg[\frac{1}{2}\, A_\mu \cdot A^\mu + \bar C \cdot C + \frac{1}{2}\, \phi \cdot \phi 
+ 2 \,\bar\beta \cdot \beta + \bar C_\mu \cdot C^\mu - \frac{1}{4}\, B^{\mu\nu} \cdot B_{\mu\nu}\bigg]. \quad
\label{LAB}
\end{eqnarray}
It is worthwhile to mention that all  
terms in the square brackets are Lorentz scalar and they are chosen in such a way that each term carries zero ghost number and mass dimension equal to two 
(in natural units: $\hbar = c = 1$) for the 4D theory. Furthermore, the 
(anti-)BRST symmetry transformations (decrease) increase the ghost number 
by one unit when they operate on any generic field.  Also, the operation of nilpotent transformations raises mass dimension by one when 
they act on any field. One can see these observations directly from the 
expressions of the (anti-)BRST symmetry transformations given in (\ref{sab1}),  (\ref{sab2}) and  (\ref{sab3}).     
The Lagrangian densities in its full blaze of glory (in the Feynman-t' Hooft gauge) can written as
\begin{eqnarray}
{\cal L}_{(B)} &=& - \frac {1}{4}\, F^{\mu\nu}\cdot F_{\mu\nu}+ \frac {1}{12}\,H^{\mu\nu\eta}\cdot H_{\mu\nu\eta}
+\frac{m}{4}\,\varepsilon_{\mu\nu\eta\kappa} B^{\mu\nu}\cdot F^{\eta\kappa} 
+  \frac{1}{2} \,\big[B \cdot B + \bar B \cdot \bar B \big] \nonumber\\
&-& B\cdot (\partial_\mu A^\mu) 
+ \big[B^\mu - g(C \times \bar C^\mu) \big]\cdot \big[B_\mu + D_\mu \phi - g(C \times \bar C_\mu) +D^\nu B_{\mu\nu} \big] \nonumber\\ 
&+& \frac{1}{2}\, \big[(D^\mu \bar C^\nu - D^\nu \bar C^\mu) - g(\bar \xi \times F^{\mu\nu}) \big] \cdot
\big[(D_\mu C_\nu - D_\nu C_\mu) - g(\xi \times F_{\mu\nu}) \big]\nonumber\\
&-& \partial^\mu \bar C \cdot D_\mu C +  D^\mu \bar \beta \cdot D_\mu \beta 
+ \frac{g}{2}\, \big[R -g (C \times \bar \xi) \big]\cdot (B^{\mu\nu} \times F_{\mu\nu}) \nonumber\\
&-& \big[\bar \lambda - g(C \times  \bar \beta)\big]\cdot \big[\rho - g(C \times  \phi) -  D_\mu C^\mu \big] 
- \big[\rho - g(C \times \phi) \big]\cdot D_\mu \bar C^\mu, 
\label{brstlag}
\end{eqnarray}
\begin{eqnarray}
{\cal L}_{(\bar B)} &=& - \frac {1}{4} \,F^{\mu\nu}\cdot F_{\mu\nu}+ \frac {1}{12}\,H^{\mu\nu\eta}\cdot H_{\mu\nu\eta}
+\frac{m}{4}\,\varepsilon_{\mu\nu\eta\kappa} B^{\mu\nu}\cdot F^{\eta\kappa} 
+  \frac{1}{2} \big[B \cdot B + \bar B \cdot \bar B \big] \nonumber\\
&+& \bar B\cdot (\partial_\mu A^\mu)
+ \big[\bar B^\mu - g(\bar C \times C^\mu)\big]\cdot \big[\bar B_\mu + D_\mu \phi - g(\bar C \times C_\mu) - D^\nu B_{\mu\nu}\big] \nonumber\\
&+& \frac{1}{2}\, \big[(D^\mu \bar C^\nu - D^\nu \bar C^\mu) - g(\bar \xi \times F^{\mu\nu})\big] \cdot
\big[(D_\mu  C_\nu - D_\nu C_\mu) - g(\xi \times F_{\mu\nu}) \big]\nonumber\\
&-& D^\mu \bar C \cdot \partial_\mu C + D^\mu \bar \beta \cdot D_\mu \beta 
- \frac{g}{2}\, \big[\bar R - g(\bar C \times \xi) \big]\cdot (B^{\mu\nu} \times F_{\mu\nu})\nonumber\\ 
&-& \big[\bar \rho - g(\bar C \times \phi) \big]\cdot \big[\lambda - g(\bar C \times  \beta) +  D_\mu C^\mu \big] 
+ \big[\lambda - g(\bar C \times \beta) \big]\cdot D_\mu \bar C^\mu.
\label{antibrstlag}
\end{eqnarray}
These are the coupled Lagrangian densities because the pairs of the NL-type auxiliary fields 
$(B, \bar B)$, $(B_\mu, \bar B_\mu)$, $(R, \bar R)$, $(\lambda, \rho)$, $(\bar \lambda, \bar \rho)$  are related to each other
through CF-type conditions (cf. (\ref{cf1}) and (\ref{cf2})). Further, the couple Lagrangian densities are equivalent  because they
respect (anti-)BRST symmetry transformations on constrained surface defined by CF-type conditions (see, Appendix {\bf C} below).

\section{(Anti-)BRST invariance of the Lagrangian densities, nilpotency and absolute anticommutativity of the (anti-)BRST symmetries: superfield approach}
It is evident form the expressions of the Lagrangian densities (\ref{LB}) and (\ref{LAB}) that the (anti-)BRST invariance can now be proven in a rather simpler 
way. This is because of the fact that under the operation of (anti-)BRST transformations, ${\cal L}$ transforms to a total spacetime derivative and 
rest  part in (\ref{LB}) and (\ref{LAB}) turns out to be zero due to the nilpotency and anticommutativity properties of the (anti-)BRST transformations.

The above  BRST and anti-BRST invariances of the coupled Lagrangian densities can also be discussed in the context of superfield formalism. 
Thus, for the sake of brevity, we generalize the Lagrangian densities on the $(4,2)$D supermanifold as
\begin{eqnarray}
\tilde {\cal L}_{(B)} &=& \tilde {\cal L} + \frac{\partial}{\partial \bar \theta} \frac{\partial}{\partial \theta} 
\bigg[\frac{1}{2}\, \tilde {\cal A}^{(h)}_\mu \cdot \tilde {\cal A}^{\mu(h)} + {\tilde {\bar {\cal F}}}^{(h)} \cdot \tilde {\cal F}^{(h)} 
+\frac{1}{2}\, \tilde \Phi^{(h)} \cdot \tilde \Phi^{(h)}
+2\, \tilde {\bar  \beta}^{(h)} \cdot \tilde \beta ^{(h)} \nonumber\\
&+& {\tilde {\bar {\cal F}}}^{(h)}_\mu \cdot \tilde {\cal F}^{\mu(h)}
 - \frac{1}{4}\, \tilde {\cal B}^{\mu\nu(h)} \cdot \tilde {\cal B}^{(h)}_{\mu\nu}\bigg], 
 \label{superL1}
\end{eqnarray}
\begin{eqnarray}
\tilde {\cal L}_{(B)} &=& \tilde {\cal L} -  \frac{\partial}{\partial \theta} \frac{\partial}{\partial \bar \theta} 
\bigg[\frac{1}{2}\, \tilde {\cal A}^{(h)}_\mu \cdot \tilde {\cal A}^{\mu(h)} + {\tilde {\bar {\cal F}}}^{(h)} \cdot \tilde {\cal F}^{(h)} 
+ \frac{1}{2}\, \tilde \Phi^{(h)} \cdot \tilde \Phi^{(h)}
+2\, \tilde {\bar  \beta}^{(h)} \cdot \tilde \beta^{(h)} \nonumber\\
&+& {\tilde {\bar {\cal F}}}^{(h)}_\mu \cdot \tilde {\cal F}^{\mu(h)}
 - \frac{1}{4}\, \tilde {\cal B}^{\mu\nu(h)} \cdot \tilde {\cal B}^{(h)}_{\mu\nu}\bigg], 
 \label{superL2}
\end{eqnarray}
where  the super-Lagrangian density $\tilde {\cal L}$ is given by
\begin{eqnarray}
\tilde {\cal L} &=& - \frac{1}{4}\, \tilde {\cal F}^{(h)}_{\mu\nu} \cdot \tilde {\cal F}^{\mu\nu(h)} 
+ \frac{1}{12}\, \tilde {\cal H}^{(h)}_{\mu \nu \eta}\cdot \tilde {\cal H}^{\mu \nu \eta (h)} 
+ \frac{m}{4}\, \varepsilon^{\mu \nu \eta \kappa}\,\tilde {\cal B}^{(h)}_{\mu \nu}\cdot \tilde {\cal F}^{(h)}_{\eta\kappa}.
\label{supertop}
\end{eqnarray}
By virtue of the HCs [cf. (\ref{hc1}) and (\ref{hc2})], the first two terms in the super-Lagrangian density $(\tilde {\cal L} )$ 
are independent of the Grassmannian 
variables $(\theta, \bar \theta)$. The key reason behind this is that these terms are gauge-invariant (and obviously (anti-)BRST invariant). 
The super-topological term in (\ref{supertop}) can be expressed, in terms of Grassmannian variables,  as
\begin{eqnarray}
\frac{m}{4}\, \varepsilon^{\mu\nu\eta\kappa} {\cal B}^{(h)}_{\mu\nu} \cdot {\cal F}^{(h)}_{\eta \kappa}
 &=& \frac{m}{4}\, \varepsilon^{\mu\nu\eta\kappa} B_{\mu\nu} \cdot F_{\eta \kappa} 
- \theta\, \partial_\mu\bigg[\frac{m}{2} \varepsilon^{\mu\nu\eta\kappa} F_{\mu\nu}\cdot \bar C_\kappa\bigg] \nonumber\\
&-& \bar \theta\, \partial_\mu\bigg[\frac{m}{2} \varepsilon^{\mu\nu\eta\kappa} F_{\mu\nu}\cdot C_\kappa\bigg] 
+ \theta \bar \theta\,  \partial_\mu\bigg[\frac{m}{2} \varepsilon^{\mu\nu\eta\kappa} F_{\mu\nu}\cdot \bar B_\kappa\bigg].
\end{eqnarray}
The (anti-)BRST invariance of the super-topological term can be captured in the context of superfield formalism as
\begin{eqnarray}
\frac{\partial}{\partial \bar \theta} \bigg[\frac{m}{4}\, \varepsilon^{\mu\nu\eta\kappa} {\cal B}^{(h)}_{\mu\nu} \cdot
 {\cal F}^{(h)}_{\eta \kappa}\bigg] \bigg|_{\theta = 0} 
&=& - \partial_\mu\bigg[\frac{m}{2} \varepsilon^{\mu\nu\eta\kappa} F_{\mu\nu}\cdot C_\kappa\bigg] 
= s_b \bigg[\frac{m}{4}\,\varepsilon^{\mu\nu\eta\kappa} B_{\mu\nu} \cdot F_{\eta \kappa} \bigg], \nonumber\\
\frac{\partial}{\partial \theta} \bigg[\frac{m}{4}\, \varepsilon^{\mu\nu\eta\kappa} {\cal B}^{(h)}_{\mu\nu} \cdot
 {\cal F}^{(h)}_{\eta \kappa}\bigg] \bigg|_{\bar \theta = 0} 
&=& - \partial_\mu\bigg[\frac{m}{2} \varepsilon^{\mu\nu\eta\kappa} F_{\mu\nu}\cdot \bar C_\kappa\bigg] 
= s_{ab} \bigg[\frac{m}{4}\,\varepsilon^{\mu\nu\eta\kappa} B_{\mu\nu} \cdot F_{\eta \kappa} \bigg],\nonumber\\
\frac{\partial}{\partial \bar \theta} \frac{\partial}{\partial \theta}\bigg[\frac{m}{4}\, \varepsilon^{\mu\nu\eta\kappa} {\cal B}^{(h)}_{\mu\nu} \cdot
 {\cal F}^{(h)}_{\eta \kappa}\bigg] &=& +  \partial_\mu\bigg[\frac{m}{2} \varepsilon^{\mu\nu\eta\kappa} F_{\mu\nu}\cdot \bar B_\kappa\bigg] 
= s_b s_{ab} \bigg[\frac{m}{4}\,\varepsilon^{\mu\nu\eta\kappa} B_{\mu\nu} \cdot F_{\eta \kappa} \bigg]. \quad
\end{eqnarray}
Thus, under the operation of Grassmannian translational generators $\partial_{\bar\theta}$,  $\partial_{\theta}$,
the super-topological term remains quasi-invariant (i.e. transforms to a total spacetime derivative). This implies that the topological term 
 remains invariant modulo a total spacetime derivative term under the operations of BRST and/or anti-BRST transformations. 
Consequently,  the super-Lagrangian densities (\ref{superL1}) and (\ref{superL2}) 
remain invariant (up to a total spacetime derivative) under the action of Grassmannian derivatives due to  the nilpotency 
(i.e. $\partial^2_{\bar\theta} = 0$,  $\partial^2_{\theta} = 0$)  and anticommutativity 
(i.e. $\partial_{\bar\theta} \partial_{\theta} + \partial_{\theta} \partial_{\bar \theta} = 0$) of the Grassmannian translation generators. 
This implies the (anti-)BRST invariance of the coupled Lagrangian densities within the framework of superfield formalism.

We can also capture the nilpotency and absolute anticommutativity properties of the (anti-)BRST symmetry transformations in the language of Grassmannian translational generators. Mathematically, 
to corroborate this statement,  the following relations are true, namely;   
\begin{eqnarray}
\frac{\partial}{\partial \bar \theta} \frac {\partial}{\partial \bar \theta}\, \tilde \Sigma^{(h)}(x, \theta, \bar \theta) = 0 
\Longleftrightarrow s^2_b  \Sigma(x) =0, \nonumber\\
\frac{\partial}{\partial \theta} \frac {\partial}{\partial \theta}\, \tilde \Sigma^{(h)}(x, \theta, \bar \theta) = 0 \Longleftrightarrow s^2_{ab}  \Sigma(x) =0,
\end{eqnarray}
\begin{eqnarray}
\bigg(\frac{\partial}{\partial \bar \theta} \frac {\partial}{\partial \theta} + 
\frac{\partial}{\partial \theta} \frac {\partial}{\partial \bar \theta}\bigg) \tilde \Sigma^{(h)}(x, \theta, \bar \theta) = 0 
\Longleftrightarrow \big(s_b\,s_{ab}+ s_{ab}\, s_b\big) \Sigma(x) = 0,
\end{eqnarray}
where $\Sigma(x)$ is any generic field present in the 4D (anti-)BRST invariant theory and 
$\tilde \Sigma^{(h)}(x, \theta, \bar \theta)$ is the corresponding superfield defined on the  $(4, 2)$D supermanifold.

\section{Conclusions}

In our present investigation, we have exploited the superfield formalism to derive the off-shell nilpotent and absolutely 
anticommuting BRST as well as anti-BRST symmetry transformations corresponding to the combined ``scalar" and ``vector" gauge 
transformations for the 4D topologically massive  non-Abelian gauge theory. In this approach, we have invoked the power and strength of 
{\it three} horizontality conditions in order to derive the complete set of the (anti-)BRST transformations. By using the basic tenets of BRST formalism, 
we have obtained the most general BRST and anti-BRST invariant Lagrangian densities  (in the Feynman gauge) for the topologically massive model 
(cf. (\ref{brstlag}) 
and (\ref{antibrstlag})), respectively, where the ghost number and mass dimension of the dynamical fields  are taken into account.

The BRST and anti-BRST invariant Lagrangian densities are coupled but equivalent due to the very existence of 
{\it five} constrained field equations defined by CF-type conditions (cf. (\ref{cf1}) and (\ref{cf2})). 
{\it Two} of them are fermionic in nature (cf. (\ref{cf2})). These CF 
conditions provide us the relations between the pairs of NL-type 
auxiliary fields. All CF-type conditions play very important role:  
\begin{enumerate}
\item in the proof of anticommutativity (i.e. linear independence) of the BRST and anti-BRST transformations (cf. (\ref{anticom1}) and (\ref{anticom2})), and 
\item in the derivation of coupled (but equivalent) Lagrangian densities.
\end{enumerate}
These CF conditions are (anti-)BRST invariant  and, thus, they are physical restrictions on the (anti-)BRST invariant theory.

  We have provided the geometrical origin of the BRST and anti-BRST symmetry transformations in the language of Grassmannian translational 
generators $\partial_{\bar \theta}$ and $\partial_\theta$, respectively. The properties of the (anti-)BRST transformations are also captured 
in terms of the Grassmannian translational generators. Further, by exploiting the key properties of Grassmannian translation generators,  
	we have also captured the (anti-)BRST invariance of the coupled Lagrangian
densities within the framework of superfield formalism in a simple and straightforward manner.          

We have observed that the vector gauge symmetry of the Kalb-Ramond field $B^a_{\mu\nu}$ 
in the non-Abelian generalization of the topologically model exists due to the introduction of an auxiliary vector field $K^a_\mu$ in the 
Lagrangian density (\ref{nonablag}) with the expression of the field strength given in Eq. (\ref{cur2}). From the (anti-)BRST 
transformations of $K^a_\mu$ as given in Eq. (\ref{sab3}):
\begin{eqnarray*}
&& s_bK^a_\mu= (D_\mu \xi)^a - C^a_\mu - g(K_\mu \times C)^a, \nonumber\\
&& s_{ab}K^a_\mu = (D_\mu \bar \xi)^a - {\bar C}^a_\mu - g(K_\mu \times \bar C)^a,
\end{eqnarray*}  
we observe that all the modes of the auxiliary field are unphysical.

We have not included matter fields in this gauge theory. Fermions can be introduced in the model via the 
coupling $\bar{\psi}\sigma^{\mu\nu}\psi B_{\mu\nu}$ where $\sigma^{\mu\nu}=\dfrac{i}{4}[\gamma^\mu,\gamma^\nu]$.  
This coupling is invariant under $CP$ transformation and remains invariant under the gauge transformations
\begin{eqnarray}
A_\mu &\to&  U A_\mu U^\dagger+\frac{i}{g}(\partial^\mu U) U^\dagger, \nonumber\\
B_{\mu\nu} &\to & U B_{\mu\nu} U^\dagger,\qquad \psi\to U\psi, \qquad \bar \psi \to  \bar \psi U^\dagger,
\end{eqnarray}
but the interaction term does not obey the vector gauge symmetry of $B_{\mu\nu}$ field. It will be interesting to see how the interaction $\bar{\psi}\sigma^{\mu\nu}\psi B_{\mu\nu}$ contribute to the chromomagnetic moment and mass renormalization of quarks in QCD. We can also think of modification of the interaction as $\bar{\psi}\sigma^{\mu\nu}\psi\big[ B_{\mu\nu}- (D_\mu K_\nu - D_\nu K_\mu) \big]$ to get an interaction term remained invariant under the vector gauge symmetry of $B^{\mu\nu}$ field. In the both cases, we should see how those interactions contribute to the beta function in the non-Abelian gauge theory because the interaction terms are new with respect to existing literature . We do not know from our present knowledge how the chiral symmetry of fermion field can be broken in this model. It should also be part in the investigation how mass of gluon is renormalized in this topologically massive model.

\section*{Acknowledgements}
RK would like to thank UGC, Government of India, New Delhi, for financial support under the PDFSS scheme.

\renewcommand{\theequation}{A.\arabic{equation}}    
\setcounter{equation}{0}  


\section*{Appendix A: Determination of various secondary fields} 
Exploiting the first HC (\ref{hc1}) for the superfields (\ref{super-ex1}), we obtain the values of secondary fields in terms of the
dynamical and auxiliary fields of the 4D (anti-)BRST invariant theory. Theses are listed as follows:     
\begin{eqnarray}
&& R_\mu = D_\mu C,\; \quad \bar R_\mu = D_\mu \bar C,\; \quad B_1 = \frac{g}{2}\, (C \times C),\;
\quad  s = - g (\bar B_1 \times C), \nonumber\\
&& \bar B_2 = \frac{g}{2}\, (\bar C \times \bar C),\; \quad
\bar B_1 + B_2 -   g(C \times \bar C) = 0,\; \quad \bar s = g (B_2 \times \bar C),\nonumber\\
&& S_\mu = D_\mu B_2 - g (D_\mu C \times \bar C) \equiv - D_\mu \bar B_1 + g (C \times D_\mu \bar C). 
\label{sec1}
\end{eqnarray}
The second relation in the third line of the above equation is nothing but the 
well-known CF condition. It is the {\it hallmark} of non-Abelian 1-form gauge theory and emerged 
very naturally within the framework of superfield approach to BRST formalism.  

Using the second HC (\ref{hc2}) together with (\ref{sup1}) and (\ref{sup2}), we obtain the values for the secondary fields
for the expansions of superfields (\ref{sup3}), namely; 
\begin{eqnarray}
\bar f_3 &=& g(\bar C \times \bar \beta), \qquad R_1 = - \beta -  g(C \times  \xi), \qquad \bar R_2 = - \bar \beta +g (\bar C \times \bar \xi),  \nonumber\\
f_2 &=& g(C \times \beta), \qquad  \bar f_1 +  f_3 - g(\bar C \times \phi) - g(C \times \bar \beta) = 0, \nonumber\\
S_1 &=& g(C \times \bar R_1) - g(\bar B \times \xi) + \bar f_2, \qquad f_1 + \bar f_2 - g(C \times \phi) - g(\bar C \times \beta) = 0, \nonumber\\
S_2 &=& g(\bar C \times R_2) - g(B \times \bar \xi) -  f_3,  
\qquad \bar b^{(1)}_\mu + b^{(2)}_\mu + D_\mu \phi - g(\bar C \times C_\mu) - g(C \times \bar C_\mu) =0, \nonumber\\
b_1 &=& g(C \times \bar f_1) - g(\bar B \times \phi) + g(\bar C \times \bar f_2) - g^2\big(\bar C \times (\bar C \times \beta) \big)\nonumber\\
&\equiv&  -g(\bar C \times f_1) + g(B \times \phi) - g(C \times f_3) + g^2\big(C \times (C \times \bar \beta) \big)\nonumber\\
b_2 &=& g(C \times \bar f_2)  - g(\bar B \times \beta) 
\equiv g(B \times \beta) -g(C \times  f_1) -g  (\bar C \times f_2)  + g^2 \big(C \times (C \times \phi) \big), \nonumber\\
b_3 &=& g(B \times \bar \beta) - g (\bar C \times f_3) 
\equiv g (\bar C \times \bar f_1) +g (C \times \bar f_3) - g (\bar B \times \bar \beta) -g^2 \big(\bar C \times (\bar C \times \phi)\big),\nonumber\\
b^{(1)}_\mu &=& - D_\mu \beta + g(C \times C_\mu), \qquad \bar b^{(2)}_\mu = - D_\mu \bar \beta + g(\bar C \times \bar C_\mu),\nonumber\\
q_\mu &=&  D_\mu \bar f_2 - g(D_\mu  \bar C \times \beta) + g(C\times \bar b^{(1)}_\mu) - g(\bar B \times C_\mu),\nonumber\\
\bar q_\mu &=& - D_\mu f_3 + g(D_\mu C \times \bar \beta) - g(\bar C \times b^{(2)}_\mu) + g(B \times \bar C_\mu),\nonumber\\
R_{\mu\nu} &=& - (D_\mu C_\nu - D_\nu C_\mu) + g(C \times B_{\mu\nu}) + g(\xi \times F_{\mu\nu}), \nonumber\\
\bar R_{\mu\nu} &=& - (D_\mu \bar C_\nu - D_\nu \bar C_\mu) + g(\bar C \times B_{\mu\nu}) + g(\bar \xi \times F_{\mu\nu}),\nonumber\\
S_{\mu\nu} &=&  g(B \times B_{\mu\nu}) - (D_\mu B_\nu - D_\nu B_\mu) + g(D_\mu C \times  \bar C_\nu) - g(D_\nu C \times \bar C_\mu) 
+ g(R_2 \times F_{\mu\nu}) \nonumber\\
&+&g^2 \big(\bar \xi \times (F_{\mu\nu} \times C)\big)  + g \bar C \times \big((D_\mu C_\nu - D_\nu C_\mu) 
- g(C \times B_{\mu\nu}) - g(\xi \times F_{\mu\nu})\big)\nonumber\\
&\equiv& (D_\mu \bar B_\nu - D_\nu \bar B_\mu) - g(D_\mu \bar C \times  C_\nu) + g(D_\nu \bar C \times C_\mu) - g(\bar B \times B_{\mu\nu})
- g(\bar R_1 \times F_{\mu\nu})\nonumber\\
&-& g^2 \big(\xi \times (F_{\mu\nu} \times \bar C)\big) - g C \times \big((D_\mu C_\nu - D_\nu C_\mu) - g(C \times B_{\mu\nu}) - g(\xi \times F_{\mu\nu})\big).
\label{sec2}
\end{eqnarray}
It is to be noted that the third and seventh  equations in the above are the CF-type conditions.  
These constrained field equations  emerge naturally when we set the coefficients of the wedge products $(d \theta \wedge d \bar \theta \wedge d \bar \theta)$,
$(d \theta \wedge d\theta \wedge d \bar \theta)$ and  $(dx^\mu \wedge d \theta \wedge d\theta)$ equal to zero due to the HC (\ref{hc2}).

Similarly, the third HC (\ref{hc3}) produces the precise values of the remaining secondary fields as 
\begin{eqnarray}
&&{\hspace{-.5cm}} P_\mu = D_\mu \xi - C_\mu - g(K_\mu \times C), \quad \bar P_\mu = D_\mu \bar \xi - \bar C_\mu - g(K_\mu \times \bar C), \nonumber\\
&& {\hspace{-.5cm}}\bar R + R + \phi - g(\bar C \times \xi) - g(C \times \bar \xi) = 0, \nonumber\\
&&{\hspace{-.5cm}} Q_\mu = D_\mu R - g(D_\mu C \times \bar \xi) + g (B \times K_\mu) - B_\mu -g \big(\bar C \times(D_\mu \xi - C_\mu - g(K_\mu \times C))\big) \nonumber\\ 
&&{\hspace{-.5cm}}~~~~\equiv - D_\mu \bar R + g(D_\mu \bar C \times \xi) - g(\bar B \times K_\mu) + \bar B_\mu 
+ g \big(C \times (D_\mu \bar \xi - \bar C_\mu - g(K_\mu \times \bar C))\big). \qquad \quad
\label{sec3}
\end{eqnarray}
The field equation in the third line of the above equation is also the CF-type condition  and it emerges naturally 
from the coefficient of Grassmannian differentials $(d\theta \wedge d\bar \theta)$ in equation (\ref{hc3}).


\renewcommand{\theequation}{B.\arabic{equation}}    
\setcounter{equation}{0}  

\section*{Appendix B: Absolute anticommutativity property of (anti-)BRST transformations} 
It is well-known that BRST and anti-BRST transformations by construction are off-shell nilpotent and absolutely anticommuting. 
The latter property is satisfied due to the existence of five CF-type conditions.  The anticommutator of the BRST and anti-BRST transformations for the 
gauge field $A_\mu$ can be written as
\begin{eqnarray}
\{s_b, \, s_{ab} \}A_\mu = D_\mu \big[B + \bar B - g \big(C \times \bar C\big)\big].
\label{anticom1}
\end{eqnarray}
Thus, it is clear that $\{s_b, \, s_{ab} \}A_\mu = 0$ on the constraint hypersurface defined by CF condition: $B + \bar B - g \big(C \times \bar C\big) = 0$. 
Similarly, for the sake of completeness,  we note that the followings are true: 
\begin{eqnarray}
\{s_b,\; s_{ab}\}B_{\mu\nu} &=& g[B + \bar B - g(C \times \bar C)]\times B_{\mu\nu}
+ g[R + \bar R -g(C \times \bar \xi) - g(\bar C \times \xi)]\times F_{\mu\nu}\nonumber\\
&-&D_\mu [B_\nu + \bar B_\nu - g(\bar C \times C_\nu) - g(\bar C \times \bar C_\nu)] \nonumber\\
&+& D_\nu [B_\mu + \bar B_\mu - g(\bar C \times C_\mu) - g(\bar C \times \bar C_\mu)],\nonumber\\
\{s_b,\; s_{ab}\} C_\mu &=& g[B + \bar B - g(C \times \bar C)] \times C_\mu 
- D_\mu [\rho + \lambda - g(C \times \phi) - g(\bar C \times  \beta)], \nonumber\\
\{s_b,\; s_{ab}\} \bar C_\mu &=& g[B + \bar B -g (C \times \bar C)] \times \bar C_\mu
- D_\mu [\bar \rho + \bar \lambda - g(\bar C \times \phi) - g(C \times \bar \beta)], \nonumber\\
\{s_b,\; s_{ab}\} \rho &=& g\big[B + \bar B - g(C \times \bar C)\big]\times \big(\rho -g(C \times \phi)\big)\nonumber\\ 
&-& g^2\big(\big[B + \bar B - g(C \times \bar C)\big]\times \phi \big) \times C,\nonumber\\
\{s_b,\; s_{ab}\} \bar \rho &=& g\big[B + \bar B - g(C \times \bar C)\big]\times \big(\bar \rho - g(\bar C \times \phi) \big)\nonumber\\ 
&-& g^2\big(\big[B + \bar B - g(C \times \bar C)\big]\times \phi \big) \times \bar C,\nonumber\\
\{s_b,\; s_{ab}\} \phi &=& g \big[B + \bar B + i (C \times \bar C)\big]\times \phi,\nonumber\\
\{s_b,\; s_{ab}\} B_\mu &=& g \big[B + \bar B -g(C \times \bar C)\big] \times B_\mu 
- g D_\mu \big[\bar \rho + \bar \lambda - g(C \times \bar \beta) - g(\bar C \times \phi) \big]\times C \nonumber\\
&+&  g^2 \big(\big[B + \bar B - g(C \times \bar C)\big] \times C\big) \times \bar C_\mu, \nonumber\\
\{s_b,\; s_{ab}\} \bar B_\mu &=& g \big[B + \bar B -g (C \times \bar C)\big] \times \bar B_\mu 
- g D_\mu \big[\rho + \lambda - g(\bar C \times \beta) - g(C \times \phi) \big]\times \bar C \nonumber\\
&+&  g^2 \big(\big[B + \bar B -g (C \times \bar C)\big] \times \bar C\big) \times  C_\mu, \nonumber\\
\{s_b,\; s_{ab}\} K_\mu &=& g \big[B + \bar B - g(C \times \bar C)\big] \times K_\mu 
+ D_\mu \big[R + \bar R + \phi - g(C \times \bar \xi) - g(\bar C \times \xi) \big] \nonumber\\
&-& \big[B_\mu + \bar B_\mu + D_\mu \phi - g(C \times \bar C_\mu) - g(\bar C \times C_\mu)\big].
\label{anticom2}
\end{eqnarray}
Thus, the anticommutativity property  of the BRST and anti-BRST transformations for the fields $A_\mu$, $B_{\mu\nu}$, $C_\mu$,  $\bar C_\mu$, $\rho$, $\bar \rho$, 
$\phi$, $B_\mu$, $\bar B_\mu$, $K_\mu$ is 
satisfied only on the constrained hypersurface defined by the CF-type conditions. For remaining  fields 
(i.e. $\beta, \bar\beta, \xi, \bar \xi, \lambda, \bar \lambda, R, \bar R$), this property
is trivially satisfied.  We again emphasize that all five CF-type conditions play an important role in providing the anticommutativity of the (anti-)BRST 
transformations and also responsible for the coupled (but equivalent) Lagrangian densities.


\renewcommand{\theequation}{C.\arabic{equation}}    
\setcounter{equation}{0}  

\section*{Appendix C: (Anti-)BRST invariance of coupled Lagrangian densities}
The Lagrangian densities ${\cal L}_{(B)}$ respects the BRST symmetry transformations, as one check that it remains quasi-invariant. To be more precise, 
${\cal L}_{(B)}$ transforms to a total spacetime derivative under the BRST transformations as follows  
\begin{eqnarray}
s_b {\cal L}_{(B)} &=& - \partial_\mu \Big[B \cdot D^\mu C 
+ \frac{m}{2}\, \varepsilon^{\mu\nu\eta\kappa} \, F_{\nu\eta}\cdot C_\kappa 
- \big(B^\mu - g (C \times \bar C^\mu)\big) \cdot \big(\rho - g(C \times \phi)\big) \nonumber\\
&-& \big(B_\nu - g(C \times \bar C_\nu) \big) \cdot \big(D^\mu C^\nu - D^\nu C^\mu 
- g (\xi \times F^{\mu\nu})\big) \nonumber\\
& -& \big(\bar \lambda - g(C \times \bar \beta)\big) \cdot D^\mu \beta \Big].
\end{eqnarray}
As a consequence, the action integral remains invariant (i.e. $s_b \int d^4x {\cal L}_{(B)} = 0$) due to Gauss divergence theorem. 
It is interesting to note that under the anti-BRST symmetry transformations ${\cal L}_{(B)}$ transforms to a total spacetime derivative plus some additional terms 
\begin{eqnarray}
s_{ab} {\cal L}_{B} &=&  - \partial_\mu \Big[\frac{m}{2}\, \varepsilon^{\mu\nu\eta\kappa} \, F_{\nu\eta}\cdot \bar C_\kappa 
- \big(B_\nu - g(C \times \bar C_\nu) \big) \cdot \big(D^\mu \bar C^\nu - D^\nu \bar C^\mu - g (\bar \xi \times F^{\mu\nu})\big) \nonumber\\ 
&+& B \cdot \partial^\mu \bar C + \big(\bar \rho - g(\bar C \times \phi)\big) \cdot \Big(B^\mu + D^\mu \phi - g (C \times \bar C^\mu) 
+ D_\nu B^{\mu\nu} \Big) \nonumber\\
&-& \big(\lambda - g(\bar C \times \beta)\big) \cdot D^\mu \bar \beta\Big] \nonumber\\
&+& D_\mu \big[\big(B + \bar B - g (C \times \bar C)\big) \cdot \partial^\mu \bar C\big] 
- \frac{g^2}{2} \big[\big(B + \bar B -g(C \times \bar C) \big) \times \bar \xi\big]  \cdot (B^{\mu\nu} \times F^{\mu\nu}) \nonumber\\
&+& g \big[\big(B + \bar B - g(C \times \bar C)\big) \times \bar \beta \big] \cdot \big[\rho - g(C \times \phi) - D_\mu C^\mu\big] \nonumber\\
&-&  \big[\bar \lambda - g (C \times \bar \beta) \big] \cdot D_\mu \big[B^\mu + \bar B^\mu + D^\mu \phi 
- g(C \times \bar C^\mu) - g(\bar C \times C^\mu) \big] \nonumber\\
&-& D_\mu \big[B_\nu + \bar B_\nu + D_\nu \phi - g(C \times \bar C_\nu) 
- g(\bar C \times C_\nu) \big] \cdot \big[D^\mu \bar C^\nu - D^\nu \bar C^\mu - g(\bar \xi \times F^{\mu\nu}) \big] \nonumber\\
&-& \frac{g}{2}\,\big[R + \bar R + \phi - g(C \times \bar \xi) - g(\bar C \times \xi) \big] \cdot 
\big[\big(D^\mu \bar C^\nu - D^\nu \bar C^\mu - g(\bar \xi \times F^{\mu\nu}) \big) \times F_{\mu\nu}\big]\nonumber\\
&-& \big[\lambda + \rho - g(C \times \phi) - g(\bar C \times \beta)\big] \cdot D_\mu (D^\mu \bar \beta) \nonumber\\
&+& \frac{g}{2}\,\big[\bar \rho + \bar \lambda - g(\bar C \times \phi) - g(C \times \bar \beta)\big] \cdot (B^{\mu\nu} \times F^{\mu\nu}).
\end{eqnarray}
Due to the validity of CF conditions, all the extra terms, except total derivative term, vanish. 
Thus, ${\cal L}_{(B)}$  also respects the anti-BRST transformations on the constrained 
hypersurfaces  defined by CF conditions (\ref{cf1}) and (\ref{cf2}).

In a similar fashion, the anti-BRST transformations leave  ${\cal L}_{(\bar B)}$ to a total spacetime derivative    
\begin{eqnarray}
s_{ab} {\cal L}_{(\bar B)} &=&  \partial_\mu \Big[\bar B \cdot D^\mu \bar C 
- \frac{m}{2}\, \varepsilon^{\mu\nu\eta\kappa} \, F_{\nu\eta}\cdot \bar C_\kappa 
- \big(\bar B^\mu - g (\bar C \times C^\mu)\big) \cdot \big(\bar \rho - g(\bar C \times \phi)\big) \nonumber\\
&-& \big(\bar B_\nu - g(\bar C \times C_\nu) \big) \cdot \big(D^\mu \bar C^\nu - D^\nu \bar C^\mu 
- g (\bar \xi \times F^{\mu\nu})\big) \nonumber\\
&+& \big(\lambda - g(\bar C \times \beta)\big) \cdot D^\mu \bar \beta
\Big].
\end{eqnarray}
Thus, ${\cal L}_{(\bar B)}$ respects off-shell nilpotent anti-BRST symmetry transformations. It is to be noted that 
under the BRST transformations ${\cal L}_{\bar B}$ transforms in the following fashion: 
\begin{eqnarray}
s_b {\cal L}_{\bar B} &=&  - \partial_\mu \Big[\frac{m}{2}\, \varepsilon^{\mu\nu\eta\kappa} \, F_{\nu\eta}\cdot C_\kappa 
+ \big(\bar B_\nu - g(\bar C \times C_\nu) \big) \cdot \big(D^\mu C^\nu - D^\nu C^\mu - g (\xi \times F^{\mu\nu})\big) \nonumber\\ 
&-& \bar B \cdot \partial^\mu C  + \big(\rho - g(C \times \phi)\big) \cdot \Big(\bar B^\mu + D^\mu \phi - g (C \times \bar C^\mu) 
- D_\nu B^{\mu\nu} \Big)  \nonumber\\
&-& \big(\bar \lambda - g(C \times \bar \beta)\big) \cdot D^\mu \bar \beta \Big] \nonumber\\
&-& D_\mu \big[\big(B + \bar B - g (C \times \bar C)\big) \cdot \partial^\mu C\big] 
+ \frac{g^2}{2}\, \big[\big(B + \bar B -g(C \times \bar C) \big) \times \xi\big]  \cdot (B^{\mu\nu} \times F^{\mu\nu}) \nonumber\\
&-& g \big[\big(B + \bar B - g(C \times \bar C)\big) \times  \beta \big] \cdot 
\big[\bar \rho - g(\bar C \times \phi) + D_\mu \bar C^\mu\big] \nonumber\\
&-& \big[\lambda - g (\bar C \times  \beta) \big] \cdot D_\mu \big[B^\mu + \bar B^\mu + D^\mu \phi 
- g(C \times \bar C^\mu) - g(\bar C \times C^\mu) \big] \nonumber\\
&+& D_\mu \big[B_\nu + \bar B_\nu + D_\nu \phi - g(C \times \bar C_\nu) 
- g(\bar C \times C_\nu) \big] \cdot \big[D^\mu C^\nu - D^\nu C^\mu - g(\xi \times F^{\mu\nu}) \big] \nonumber\\
&+& \frac{g}{2}\,\big[R + \bar R + \phi - g(C \times \bar \xi) - g(\bar C \times \xi) \big] \cdot 
\big[\big(D^\mu C^\nu - D^\nu C^\mu - g(\xi \times F^{\mu\nu}) \big) \times F_{\mu\nu}\big]\nonumber\\
&-&  \big[\bar \lambda + \bar \rho - g(\bar C \times \phi) - g(C \times \bar \beta)\big] \cdot D_\mu (D^\mu  \beta) \nonumber\\
&-& \frac{g}{2}\,\big[\rho + \lambda - g(C \times \phi) - g(\bar C \times \beta)\big] \cdot (B^{\mu\nu} \times F^{\mu\nu}).
\end{eqnarray}
It is clear that Lagrangian density ${\cal L}_{(\bar B)}$ also respects the 
BRST symmetry transformations due to the validity of  CF-type conditions. 
As a consequence, the coupled Lagrangian densities 
respect BRST and anti-BRST symmetries on the constrained hypersurface 
defined by the  CF-type conditions. This shows that the coupled Lagrangian densities are equivalent on the constrained hypersurface.



\end{document}